\documentclass[preprint,aps,floats,nofootinbib,amssymb]{revtex4}
\usepackage{epsf,epsfig}

\usepackage{graphicx}

\newcommand{\be}{\begin{equation}}
\newcommand{\ee}{\end{equation}}
\newcommand{\bea}{\begin{eqnarray}}
\newcommand{\eea}{\end{eqnarray}}
\newcommand{\nn}{\nonumber}

\renewcommand{\check}{({\bf This statement needs to be checked!!})}

\newcommand{\lsim}{
\mathrel{\hbox{\rlap{\hbox{\lower4pt\hbox{$\sim$}}}\hbox{$<$}}}}
\newcommand{\gsim}{
\mathrel{\hbox{\rlap{\hbox{\lower4pt\hbox{$\sim$}}}\hbox{$>$}}}}

\newcommand{\half}{\frac{1}{2}}

\newcommand{\mueff}{\mu_{\rm eff}}

\newcommand{\ba}{\begin{array}}
\newcommand{\ea}{\end{array}}
\newcommand{\bi}{\begin{itemize}}
\newcommand{\ei}{\end{itemize}}

\newcommand{\N}{\chi^0}
\newcommand{\C}{\chi^\pm}

\preprint{
\hbox to \hsize{
\hfill$\vcenter{\hbox{\bf MADPH-1471}
            \hbox{\bf UPR-1160-T}
                \hbox{\bf hep-ph/0609068}
                \hbox{September 2006}}$}
}

\begin{document}

\title{\vspace*{.75in}
Neutralino Signatures of the Singlet Extended MSSM}

\author{
Vernon Barger$^1$, Paul Langacker$^{2,3}$, and Gabe Shaughnessy$^1$}

\affiliation{
$^1$Department of Physics, University of Wisconsin,
Madison, WI 53706 \\
$^2$Department of Physics and Astronomy, University of Pennsylvania,
Philadelphia, PA 19104\\
$^3$ School of Natural Sciences, Institute for Advanced Study, 
Einstein Drive, Princeton, NJ 08540
\vspace*{.5in}}

\thispagestyle{empty}

\begin{abstract}
\noindent
Extending the Higgs sector of the MSSM by the addition of a gauge singlet scalar field can remedy the $\mu$ problem.  We explore the implications of extended models for both the spectrum of the neutralinos and the cascade decays of the neutralinos and charginos.  Extra steps due to light decoupled neutralinos in the cascade decays of both neutralinos and charginos allow an excess of trilepton events compared to the MSSM and the existence of events with higher  lepton multiplicity.  Additionally, displaced vertices of the $\N_2$ due to small decay widths in some models may be observable.  
\end{abstract}
\maketitle

\section{Introduction}

Discovering the underlying physics at the TeV scale is the primary goal of the Large Hadron Collider (LHC) and the future International Linear Collider (ILC).  If supersymmetry (SUSY) is found to exist it will be important to differentiate its many variations.  One such interesting class of supersymmetric model, the singlet-extended Minimal Supersymmetric Standard Model (xMSSM), exhibits characteristics not found in the MSSM \cite{Barger:2006dh}.  

The primary motivation for extending the MSSM to include a singlet field in the superpotential is to provide a solution to the $\mu$ problem \cite{muproblem}.  The $\mu$-parameter is the mass term for the Higgsinos which enters the chargino and neutralino mass matrices.  By extending the $\mu$ parameter, the only dimensionful parameter that is SUSY invariant, to an effective parameter dynamically created by the vacuum expectation value (vev) of a singlet field, the problem of fine tuning $\mu$ to the electroweak symmetry breaking scale is resolved.  Otherwise, the $\mu$ parameter is naturally either $M_{Pl}$, the Planck mass, or zero due to some higher symmetry preventing the $\mu$ term in the Lagrangian \footnote{For an alternative solution, see \cite{Giudice:1988yz}.}.  However, a vanishing $\mu$ term is in conflict with present chargino mass limits from LEP \cite{ref:chargino}.  

The superpotential term $\lambda \hat H_u \hat H_d \hat S$ of singlet extended models gives an effective $\mu$-term
\be 
\mueff = \lambda \langle S \rangle= {\lambda s\over \sqrt 2},
\ee
where $\langle S \rangle={s\over \sqrt2}$ is the vev of the singlet field.  Depending on what symmetry is imposed on the singlet field interactions, the superpotential will include other terms associated with that symmetry that are unique to the model.  In Table \ref{tbl:models}, we list these symmetries for the Next-to-Minimal Supersymmetric Standard Model (NMSSM) \cite{NMSSM}, the Minimal Nonminimal Supersymmetric Standard Model (MNSSM) also known as the nearly Minimal Supersymmetic Standard Model (nMSSM) \cite{nMSSM,Menon:2004wv}, the $U(1)'$-extended Minimal Supersymmetric Standard Model (UMSSM) \cite{UMSSM}, and  the secluded $U(1)'$-extended Minimal Supersymmetric Standard Model (sMSSM) \cite{sMSSM,Han:2004yd}.  Table \ref{tbl:models} gives the additional superpotential term of each model and the number of states in the neutralino and neutral Higgs sectors.  The additional singlet increases the number of the neutral states of the nMSSM and NMSSM by one as compared to the MSSM.  

The UMSSM has an additional neutralino and only one CP-odd neutral Higgs boson.  The associated goldstone is eaten by the $Z'$, removing one degree of freedom from the Higgs sector.  The supersymmetric partner of the $Z'$, the $Z'$ino, is an additional neutralino.  

The sMSSM contains three secluded singlet fields, $S_i$, in addition to the one that generates a $\mu$ term.   These fields increase the number of neutral states in the neutral Higgs and neutralino sectors by three.  If the coupling among the secluded singlet fields, $\lambda_S$, in the superpotential is small and the secluded singlet vevs are large, this model can mimic the nMSSM with the states associated with the secluded singlet fields decoupled \cite{Barger:2006dh}.  Therefore, we refer to the results from the nMSSM as the n/sMSSM since the above limit of the secluded model has similar behavior to the nMSSM.  For a recent comprehensive review of supersymmetric singlet models, see Ref. \cite{Kraml:2006ga}.  

The charged Higgs and chargino sectors are the same for all the models, with one charged Higgs boson and two charginos in all models.

\begin{table}[ht]
\begin{center}
\begin{tabular}{|c|clclclclc|}
\hline
Model:& MSSM &NMSSM &nMSSM& UMSSM & sMSSM\\
\hline
Symmetry:  &--  &~~ $\mathbb Z_3$    & $\mathbb Z^R_5, \mathbb Z^R_7$      & ~~$U(1)'$&$U(1)'$ \\\hline
Extra   &--         &       ~~${\kappa\over3} \hat S^3$    & $t_F  \hat S$& ~~-- &$\lambda_S S_1 S_2 S_3$ \\
superpotential term&   --      &      (cubic)    & (tadpole) & ~~-- & (trilinear secluded)\\\hline
 $\N_i$ &        	4	  &	~~~5	&  	5	 & ~~6 & 9\\
 $H^0_i$ &       	2	  &	~~~3	&  	3	 & ~~3 & 6\\
 $A^0_i$ &        	1	  &	~~~2	&  	2	 & ~~1 & 4\\
\hline
\end{tabular}
\caption{Symmetries associated with each model and their respective terms in the superpotential; the number of states in the neutralino and Higgs sectors are also given.  All models have two charginos, $\C_i$, and one charged Higgs boson, $H^\pm$.}
\label{tbl:models}
\end{center}
\end{table}

An important search channel for SUSY at hadron colliders is trileptons arising from neutralino and chargino cascade decays.  This signal from $\N_2 \C_1$ associated production and their subsequent decays has been well studied in the context of the MSSM and constrained MSSM (CMSSM) \cite{ref:trilep}.  The trilepton signal of the singlet-extended models can be different from the MSSM as the masses and couplings of the neutralinos are modified.  In addition, five (or more) lepton signals occur due to an extra cascade to a light singlino.  We investigate the generic changes in cascade decays and the associated differences in the signals.

The remainder of this paper is organized as follows:  In Section \ref{sect:spect}, we present the neutralino spectrum of these models.  In Section \ref{sect:sig}, we discuss the neutralino and chargino cascade decays including 3 lepton, 5 lepton and 7 lepton events and displaced vertices of long lived neutralinos.  Finally, in Section \ref{sect:concl}, we provide concluding remarks and future directions.

\section{Neutralino spectra}\label{sect:spect}

In the singlet extended models considered here at least one new neutralino state beyond the MSSM exists.  Depending on the model, the neutralino states can include four MSSM-like states and one nearly decoupled singlino state, or the singlino can  significantly mix with the other states, as determined from the neutralino mass matrix
\bea
{\cal M}_{\N} = \left( \begin{array} {c c c c |c| c}
	M_1 	&	0&	{-g_1 v_d/ 2}&	{g_1 v_u / 2}&	0&  0\\
	0 	&M_2&	{g_2 v_d / 2}&	{-g_2 v_u / 2}&	0&  0\\
	{-g_1 v_d / 2} 	&	{g_2 v_d / 2}&	0&	-\mu_{\rm eff}&	-\mu_{\rm eff}v_u/s&  {g_{1'}} Q_{H_d} v_d\\
	{g_1 v_u / 2} 	&	{-g_2 v_u / 2}&	-\mu_{\rm eff}& 0&	-\mu_{\rm eff}v_d/s&  {g_{1'}} Q_{H_u} v_u\\
	\hline
	0&0&-\mu_{\rm eff} v_u/s&-\mu_{\rm eff} v_d/s&\sqrt 2 \kappa s&{g_{1'}} Q_{S} s\\
	\hline
	0&0&{g_{1'}} Q_{H_d} v_d&{g_{1'}} Q_{H_u} v_u&{g_{1'}} Q_{S} s& M_{1'}\\
	\end{array} \right).
	\label{eq:neutmass}
\eea
Here the gaugino mass terms are $M_1, M_{1'}$, and $M_2$ for the $U(1)_Y, U(1)'$, and $SU(2)_L$ gauginos, respectively.  Similarly, $g_1, g_{1'}$, and $g_2$ are the corresponding gauge couplings.   For simplicity, we assume gauge coupling and gaugino mass unification:  $g_{1'}=\sqrt{5\over3} g_1$ and $M_1=M_{1'} = {5 g_1^2\over 3 g_2^2} M_2$, respectively.  

The upper left $4\times4$ submatrix is the MSSM neutralino mass matrix.  The n/sMSSM and NMSSM mass matrices are the upper left $5\times5$ matrix, with $\kappa \to 0$ for the n/sMSSM.  In the decoupling limit of large singlet vev $s$, the distinctive features of the models are transparent.  In the NMSSM, there is a heavy singlino of mass $\sqrt 2 \kappa s$ which approximately decouples from the rest of the neutralino spectrum \footnote{Approximate analytic formulae for the neutralino masses in the NMSSM can be found in Ref. \cite{ref:chimass}.  Here, we obtain the exact masses by numerical diagonalization of Eq. (\ref{eq:neutmass})}.  In contrast, the n/sMSSM has a very light neutralino which is dominantly singlino.  

The UMSSM includes the entire matrix with $\kappa \to 0$ and contains additional terms involving the Higgs charge under the $U(1)'$ symmetry.  This symmetry can result from breaking a high scale symmetry such as $E_6\to SO(10)\times U(1)_\psi \to SU(5)\times U(1)_\chi \times U(1)_\psi$.  We assume this breaking and parameterize the Higgs charges by the mixing angle between the two surviving $U(1)_{\chi,\psi}$ symmetries, $\theta_{E_6}$ (see Eq. 8 of Ref. \cite{Barger:2006dh} for the relations of the charges to $\theta_{E_6}$.)
In the large $s$ (decoupling) limit (indicated by $Z'$ mass and $Z-Z'$ mixing limits) and $M_{1'}$ near the EW scale this model has two heavy neutralinos with masses \cite{Barger:2005hb}
\be
M_{\N_{\widetilde S,\widetilde Z'}}= \vert {M_{1'}\over 2}\pm g_{1'} Q_S s\vert.
\ee

\begin{figure}[htpb]
\begin{center}
\includegraphics[angle=-90,width=0.49\textwidth]{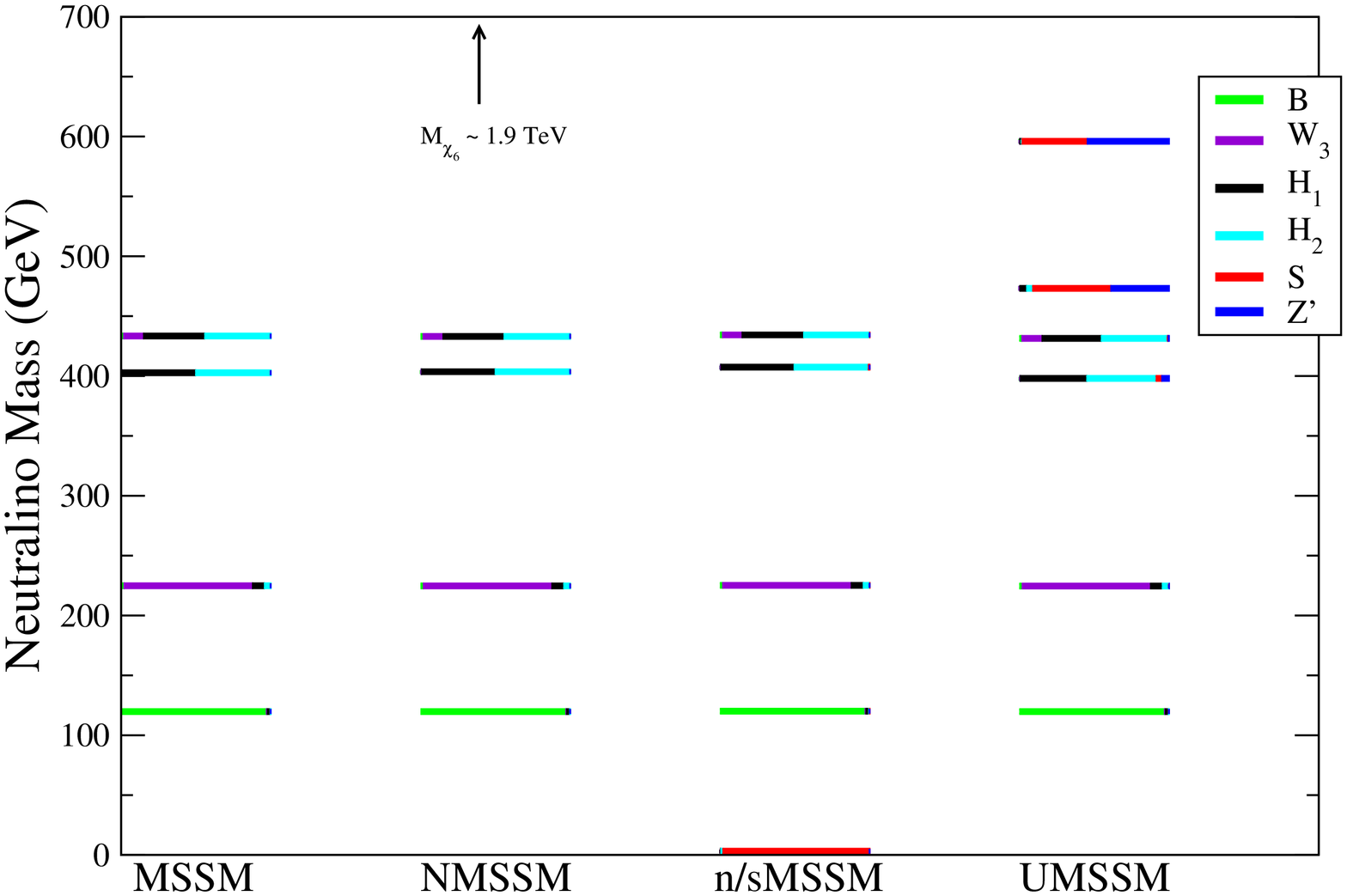}
\includegraphics[angle=-90,width=0.49\textwidth]{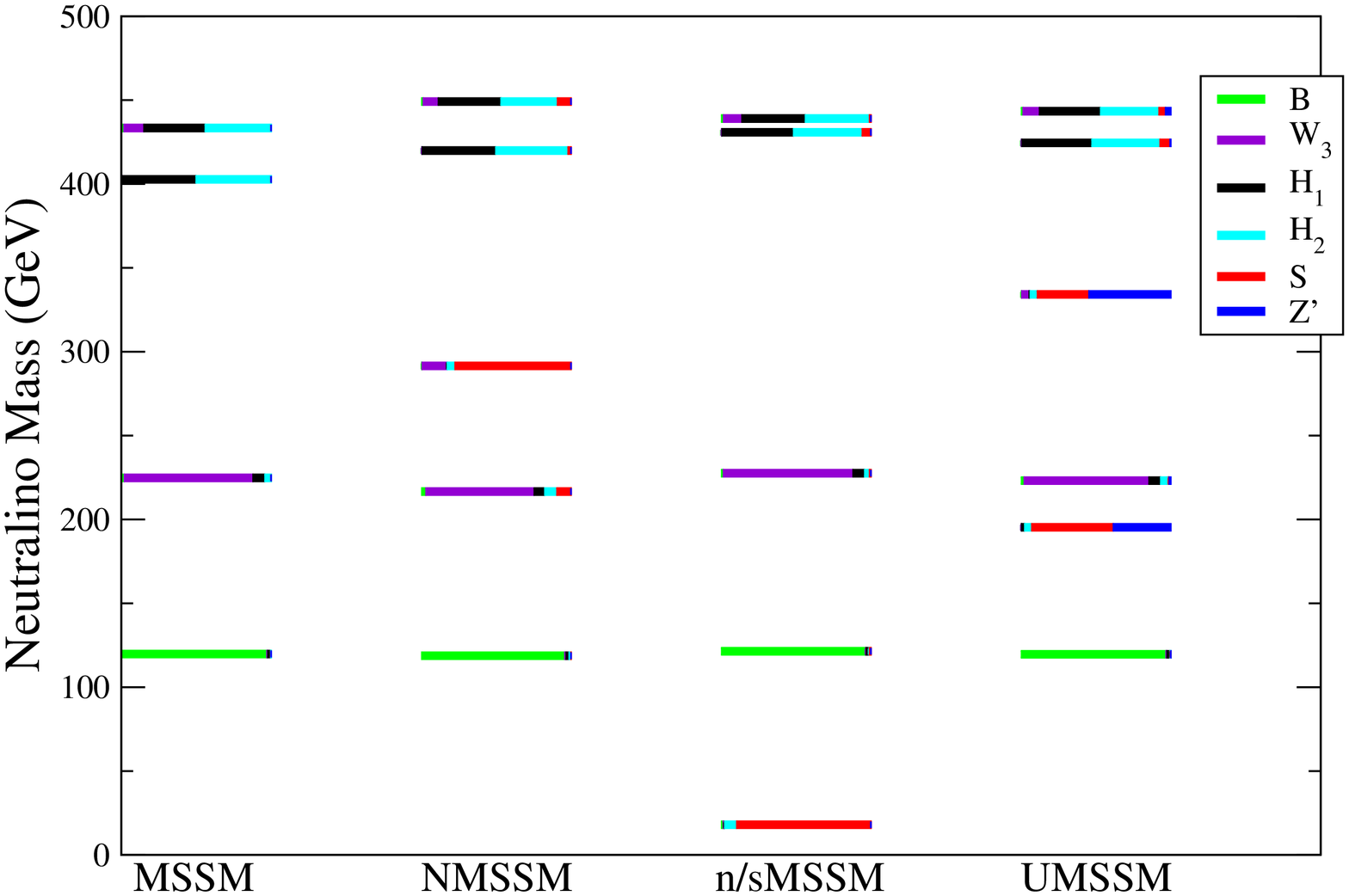}
(a)\hspace{0.48\textwidth}(b)
\caption{Illustrative neutralino composition for the models in (a) a decoupled singlino scenario and (b) a strongly mixed singino scenario.  Here, the MSSM contains a light Bino and Wino and heavy Higgsinos.  The NMSSM has a similar spectrum, but contains an additional heavy neutralino, while the n/sMSSM has a very light extra neutralino.  The UMSSM has two additional neutralinos that can intermix; their masses are strongly dependent on the singlet Higgs charge under the $U(1)'$ symmetry and the corresponding gaugino mass value.  Common parameters used for this illustration are $\tan \beta = 3$, $\mueff = 400$ GeV, $M_2 = 250$ GeV, and $\theta_{E_6} = {\pi\over 4}$ for the UMSSM.  For (a) $\kappa=0.65$ in the NMSSM and $s = 2$ TeV for the extended models; while for (b) $\kappa=0.25$ in the NMSSM, $s = 800$ GeV for the NMSSM and n/sMSSM while the UMSSM illustration has $s= 1$ TeV to satisfy the $Z-Z'$ mixing constraints \cite{Barger:2005hb}.}
\label{fig:level-light}
\end{center}
\end{figure}

The singlet models of Fig. \ref{fig:level-light}a, with a nearly decoupled singlino state, contain a neutralino sub-spectrum similar to the MSSM.  In the n/sMSSM, the singlino is lighter than the MSSM neutralino states and is thereby accessible in cascade decays.  In the NMSSM spectrum of Fig. \ref{fig:level-light}, the singlino is heavier than the other neutralino states and is therefore not likely to be produced.  

In a more general circumstance, significant mixing can occurs between the MSSM states and the singlino states, as illustrated in Fig. \ref{fig:level-light}b.  Some neutralino states can have an appreciable, but not dominant contribution of singlino and this has an impact on the decay branching fractions.  Note that in the n/sMSSM the lightest neutralino is usually still decoupled.  This will be the case unless $s\sim \min\{\mueff,v\}$ in which case the singlino terms in the neutralino mass matrix are of comparable size to the higgsino terms.

\subsection{Parameter scans}\label{sect:const}

To obtain the neutralino and chargino spectrum, we performed a random scan over the relevant parameters in the neutralino sector until $10^4$ points in parameter space were found to be consistent with the constraints listed below.  Additionally, we scanned over parameters of the Higgs sector so that we can accurately evaluate the neutralino and chargino decays involving Higgs bosons.  Parameters specific to the Higgs sector ($A_s, A_\kappa$, and $t_S$) are present in the soft Lagrangian \cite{Barger:2006dh}\footnote{Note the parameter conventions here follow Ref. \cite{Barger:2006dh} but differ slightly.  Specifically, the Higgs parameter $\lambda$ is equivalent to $h_s$ and the parameters of the n/sMSSM, $t_F$ and $t_S$ are equivalent to $M_{\rm n}^2 \xi_F$ and $M_{\rm n}^3 \xi_S$, respectively.}.  The scan parameters and ranges are summarized in Tbl. \ref{tbl:scan}.  

\begin{table}[t]
\caption{Model independent and dependent parameters and their ranges allowed in our scan.  We ignore the possibility of CP-violating phases}
\begin{center}
\begin{tabular}{|c|}
\hline
Model Independent parameters\\
\hline
	$1\le\tan \beta\le50$ \\ 
	 $-500\text{ GeV}\le M_2\le500\text{ GeV}$\\ 
	$100\text{ GeV} \le\mu_{\rm eff}\le 1\text{ GeV} $ \\
	$100\text{ GeV} \le s\le 2\text{ TeV}$ \\ 
	$0\text{ TeV}\le A_s \le 1\text{ TeV}$ \\ 
	 $-1\text{ TeV}\le A_t\le1\text{ TeV}$ \\ 
\hline
\end{tabular}
\hspace{1in}
\begin{tabular}{|c|c|c|c|c|c|}
\hline
Model dependent parameters\\
\hline
	$-0.75\le \kappa\le 0.75$ \\ 
	$-1\text{ TeV}\le A_\kappa\le 1\text{ TeV}$\\ 
	$-500\text{ GeV} \le t_S^{1/3}\le 500\text{ GeV}$  \\ 
	$-(500\text{ GeV})^2 \le t_F\le (500\text{ GeV})^2$ \\ 
	$0\le\theta_{E_6}\le \pi$ \\\hline
\end{tabular}
\end{center}
\label{tbl:scan}
\end{table}

The allowed ranges of the neutralino masses and couplings are limited by experimental constraints in both the neutralino and Higgs sectors.  For a summary of the constraints specific to the Higgs and $Z'$ sectors, see Ref. \cite{Barger:2006dh}.  There are further constraints on each model through neutralino contributions to the invisible $Z$ width and supersymmetric contributions to the anomalous magnetic moment of the muon \cite{ref:gm2}.  We found that constraints due to a $\sim 3 \sigma$ anomaly in $(g-2)_\mu$ \cite{Bennett:2004pv} are rather loose, as they can be evaded with large sfermion masses, so we did not include them.  Further constraints from the neutralino relic density $\Omega_{\N_1}$ have been computed in Ref. \cite{Barger:2005hb,Menon:2004wv}.  The n/sMSSM is constrained by the three year WMAP measurement $\Omega_{\N_1} h^2 = 0.127^{+0.007}_{-0.013}$ (where $h=0.73\pm 0.03$) \cite{Spergel:2006hy} that places a lower bound on the lightest neutralino mass of $\N_1 \gtrsim 30$ GeV.  This calculation assumed annihilation through a $Z$ boson in the s-channel.  Since other processes may increase the overall neutralino annihilation rate (or allow decays into an almost decoupled lighter neutralino in the n/sMSSM) and possibly lead to a more relaxed $m_{\N_1}$ bound, we do not enforce this lower bound on the neutralino mass.  

A very light neutralino can significantly impact the available parameter space of other MSSM parameters.  The requirement that the LSP be neutral removes a portion of MSSM parameter space where the stau is lighter than $\N_1$.  However, in the n/sMSSM, this constraint is not as limiting because the $\N_1$ can be very light.

\begin{figure}[htbp]
\begin{center}
\includegraphics[angle=-90,width=0.49\textwidth]{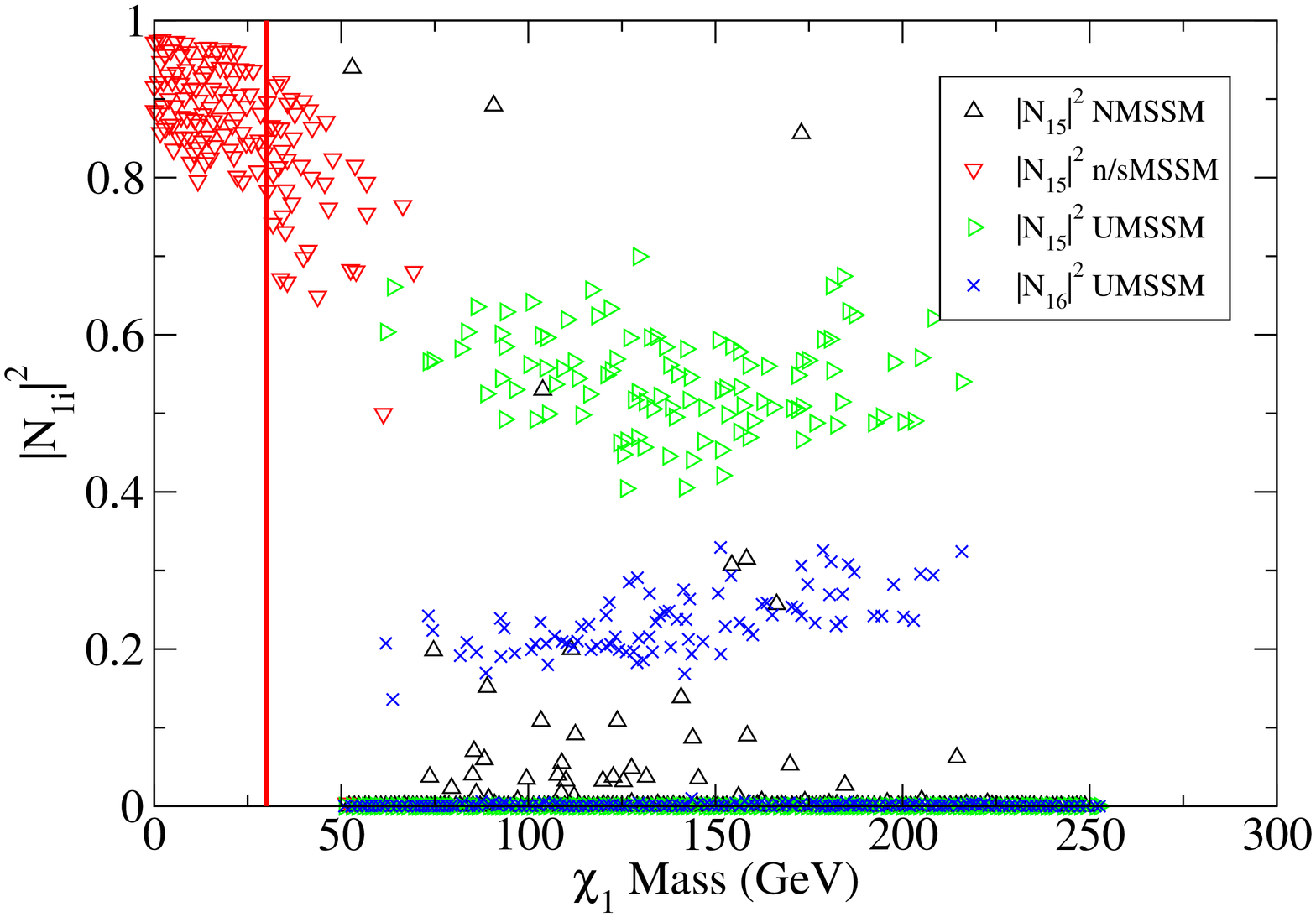}
\includegraphics[angle=-90,width=0.49\textwidth]{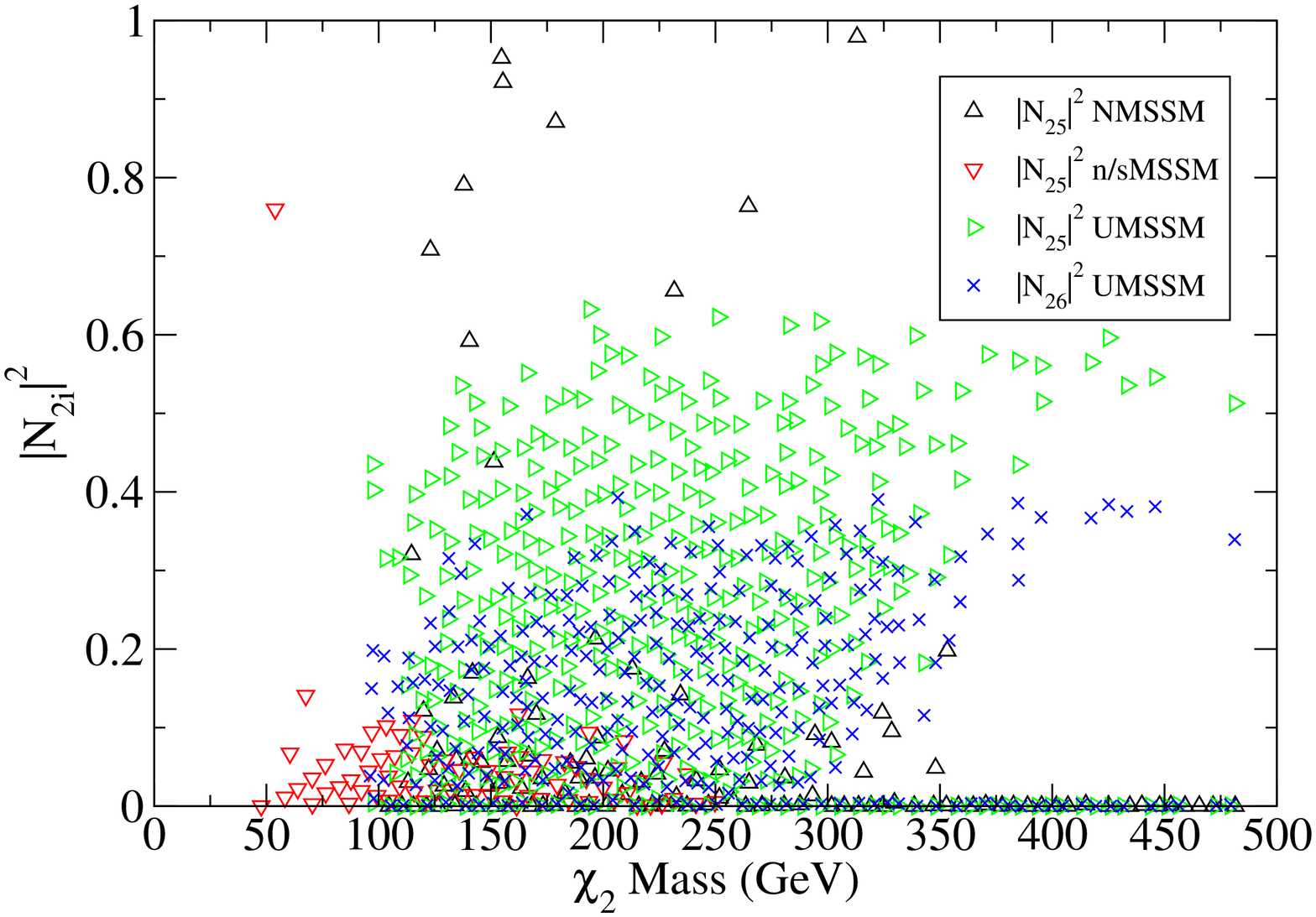}
(a)\hspace{0.48\textwidth}(b)
\includegraphics[angle=-90,width=0.49\textwidth]{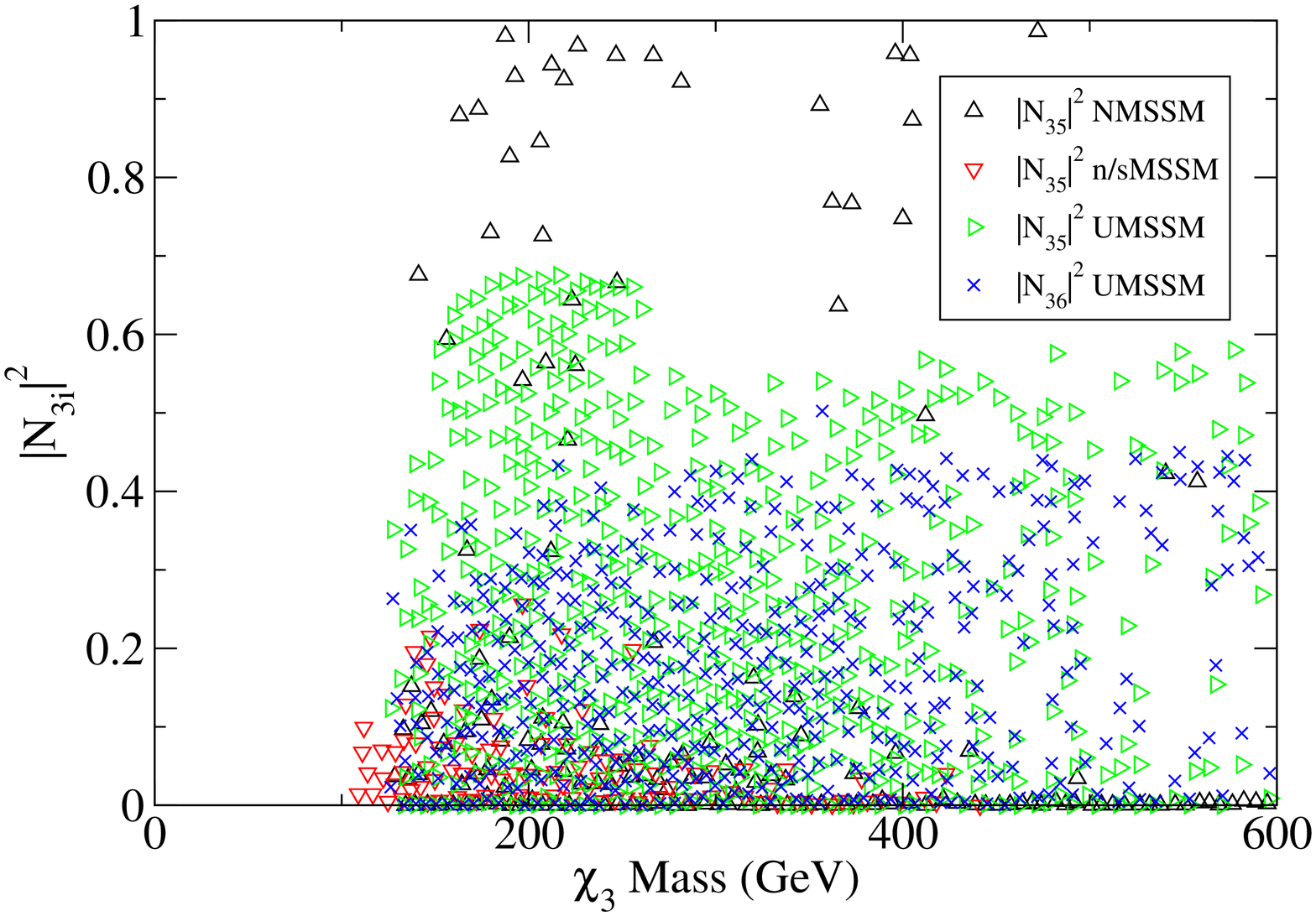}
\includegraphics[angle=-90,width=0.49\textwidth]{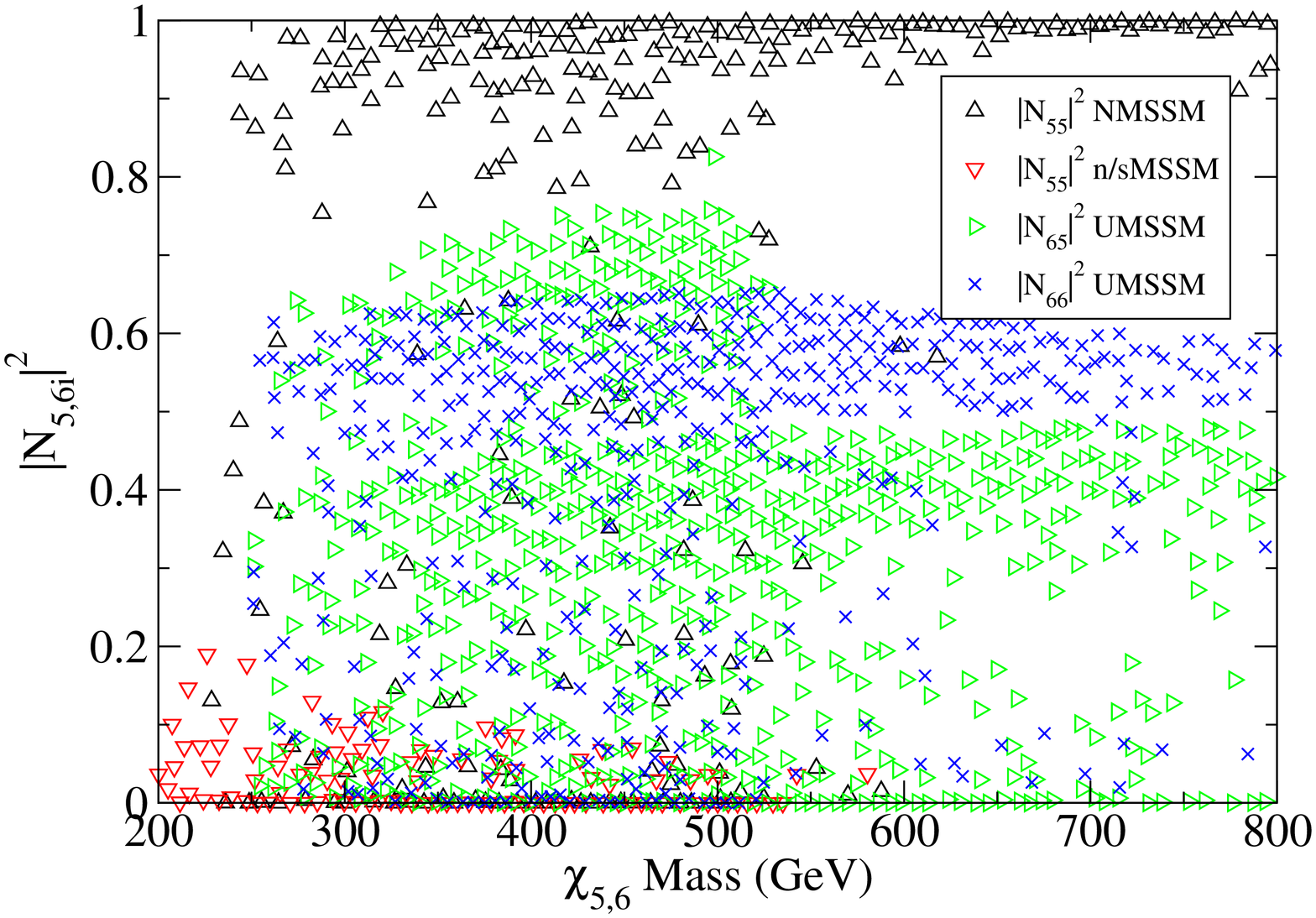}
(c)\hspace{0.48\textwidth}(d)
\caption{Singlino and $Z'$ino composition of (a-c) the three lightest neutralinos and (d) the heaviest neutralino.  The singlino composition is denoted with a black up triangle for the NMSSM, a red down triangle for the n/sMSSM and a green left triangle for the UMSSM, while the $Z'$ ino contribution is denoted with a blue x.  The lightest neutralino is dominantly singlino in the n/sMSSM, while the UMSSM's heavier neutralinos are naturally singlino and $Z'$-ino.  In the NMSSM, most of the parameter range yields a heavy neutralino that is dominantly singlino.  The vertical line in (a) is the loose $\N_1$ lower mass bound via constraints from $\Omega_{\N_1} h^2$.}
\label{fig:singcomp}
\end{center}
\end{figure}

The singlino, $|N_{i5}|^2$, and $Z'$-ino, $|N_{i6}|^2$, compositions of the light neutralinos are shown in Fig. \ref{fig:singcomp}a where $i$ is the mass index.  The NMSSM and UMSSM can have a dominantly singlino lightest neutralino (as can the $Z'$-ino in the UMSSM), but it is less natural to achieve this compared to the n/sMSSM.  The second lightest neutralino in the UMSSM, however, can have a significant fraction of singlino and $Z'$-ino states as seen in Fig. \ref{fig:singcomp}b.  We also see that the heaviest state in the NMSSM is often dominantly singlino (or $Z'$-ino in the UMSSM).  This is expected since most of the allowed parameter space in these models is in the $s$ decoupling limit, yielding a heavy, decoupled neutralino.  

\section{Multilepton Cascade Decay Signals}\label{sect:sig}

Due to the additional neutralino states, new cascade decay chains of the neutralinos and charginos can be realized in singlet extended models.  In the following, we examine the decays of neutralinos and discuss their differences from those of the MSSM as a means of experimentally identifying models.

We calculate the partial decay widths with the program SDECAY \cite{Muhlleitner:2004mk} after appropriately modifying neutralino couplings to include the effects of new particles as given in Appendix \ref{apx:couplings}.  We focus on the lighter neutralino and chargino decay modes as these states should be produced more copiously at the LHC and at a future ILC.  

\subsection{Neutralino Decays}

The neutralinos can decay via the following two body processes
\be
\N_i \to \N_j H_k,~  \N_j Z,~  H^\mp \C_j,~  W^\mp \C_j,
\label{eq:2bdyNdec}
\ee
which are the dominant decay modes for the neutralinos when kinematically allowed.  These modes are the focus of our discussion.  Otherwise, three body decays $\N_i \to \N_j f \bar f$ become relevant.  If the $Z$ decays are leptonic, the flavor subtracted cross section, $\sigma_{\N_i\to\N_j e^+ e^-}+\sigma_{\N_i\to\N_j \mu^+ \mu^-}-\sigma_{\N_i\to\N_j e^+ \mu^-}-\sigma_{\N_i\to\N_j e^- \mu^+}$, can substantially reduce the SM background to statistical fluctuations, with the SUSY signal remaining \cite{ref:kinedge-lhc}.  We treat the squarks and sleptons in the neutralino decay chains as off-shell for simplicity and to emphasize the new features in these models from the on shell decays in Eq. \ref{eq:2bdyNdec}.  

\begin{figure}[htbp]
\begin{center}
\includegraphics[angle=-90,width=0.49\textwidth]{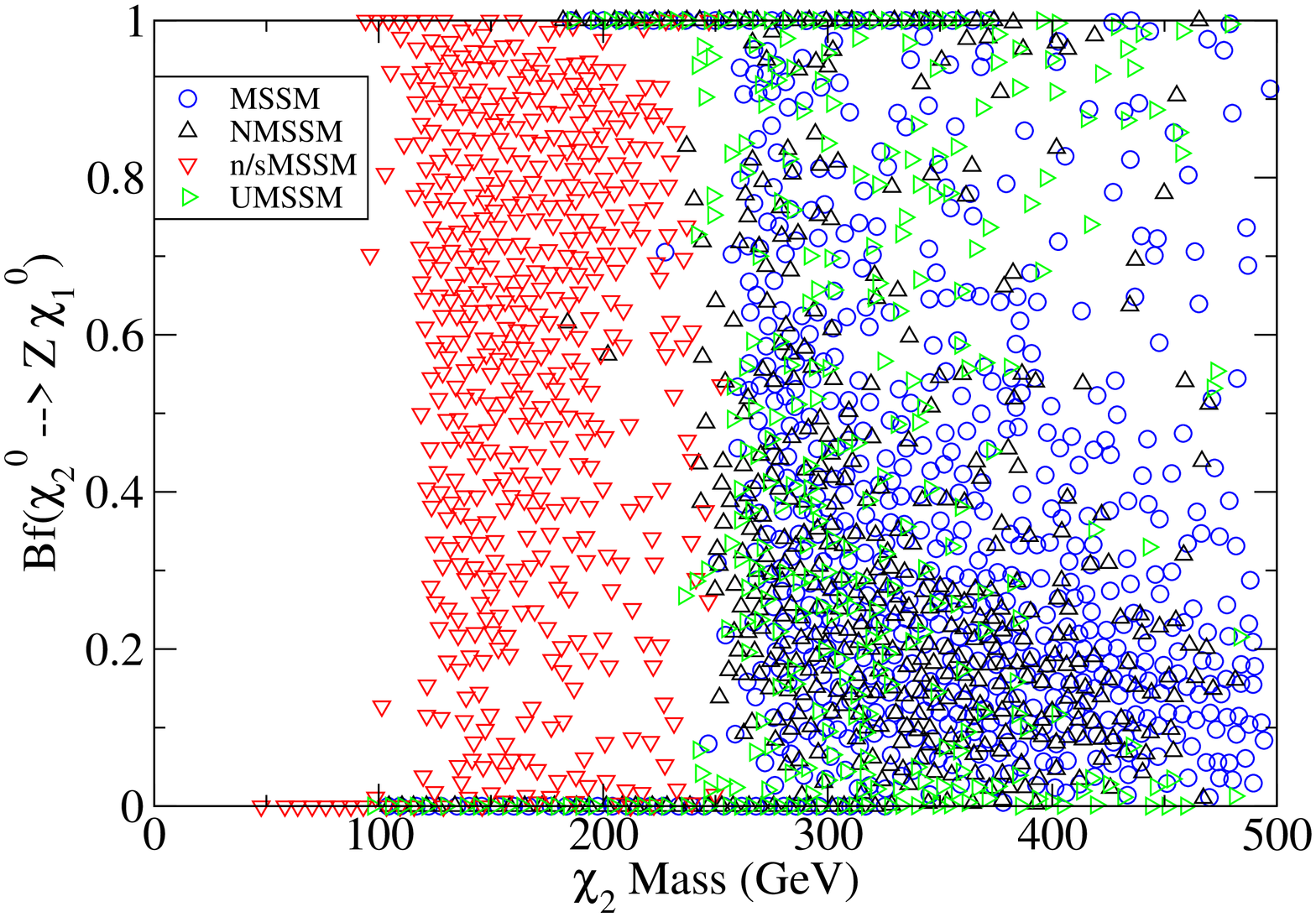}
\includegraphics[angle=-90,width=0.49\textwidth]{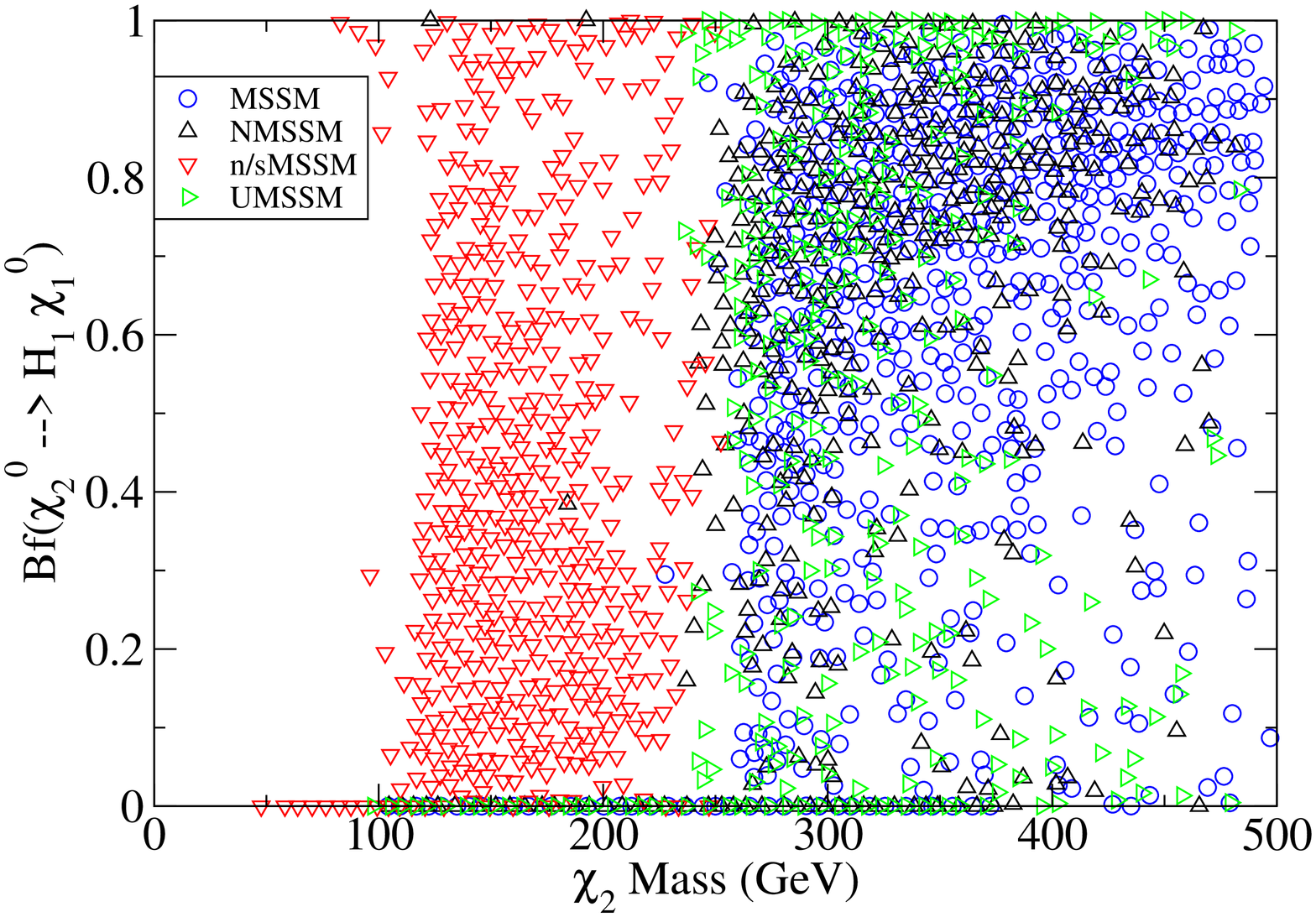}
(a)\hspace{0.48\textwidth}(b)\vspace{-.25in}
\includegraphics[angle=-90,width=0.49\textwidth]{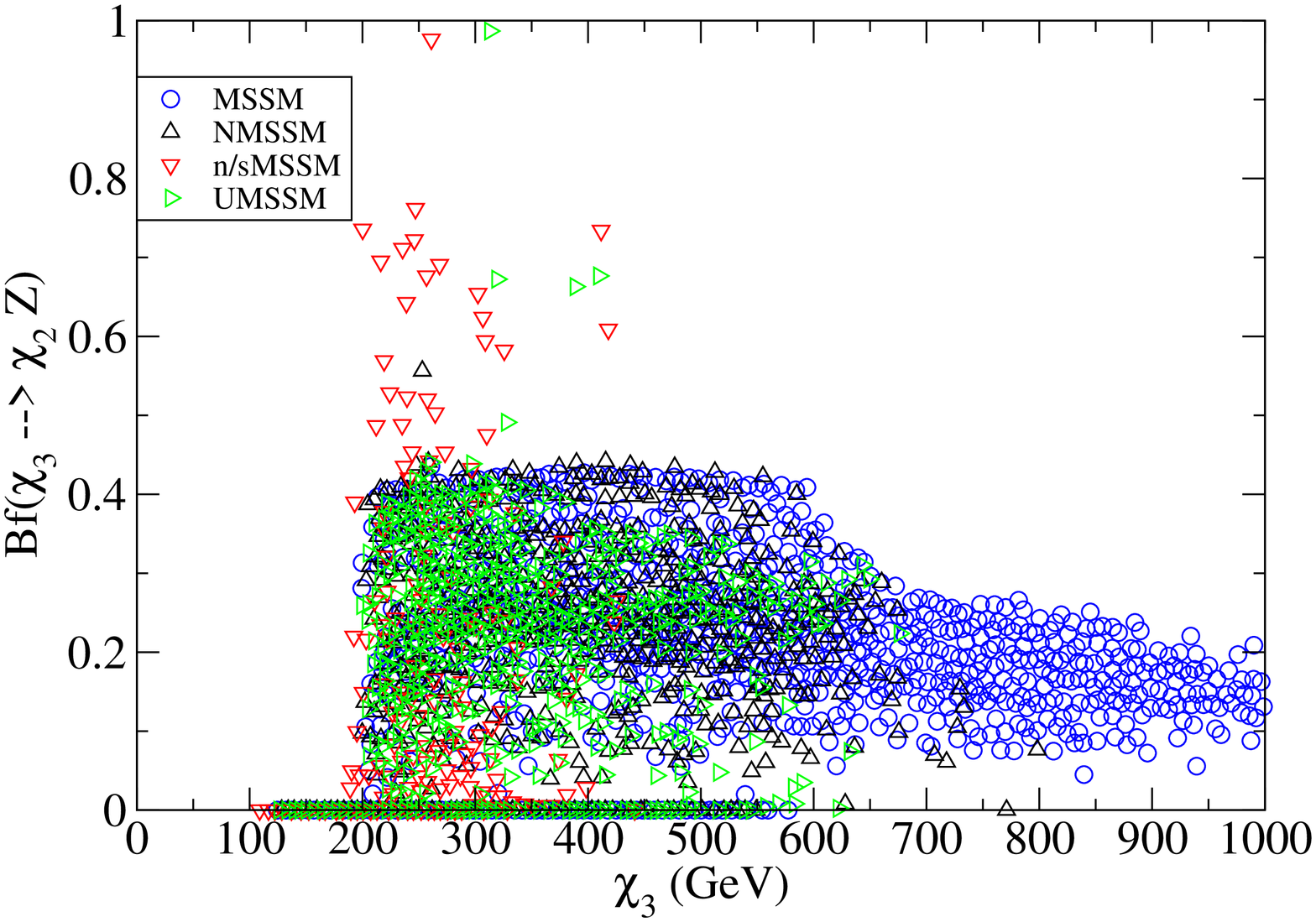}
\includegraphics[angle=-90,width=0.49\textwidth]{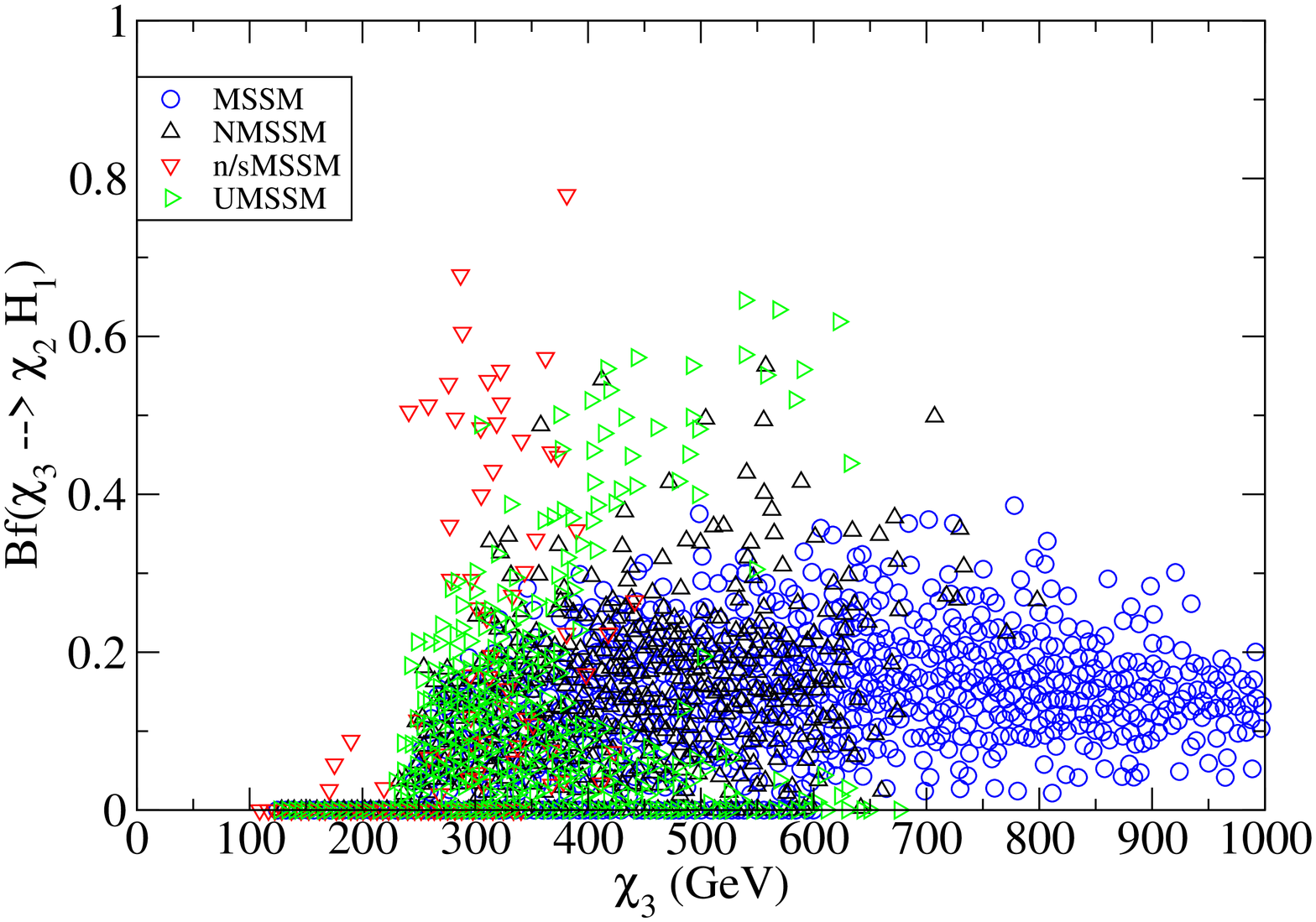}
(c)\hspace{0.48\textwidth}(d)\vspace{-.25in}
\includegraphics[angle=-90,width=0.49\textwidth]{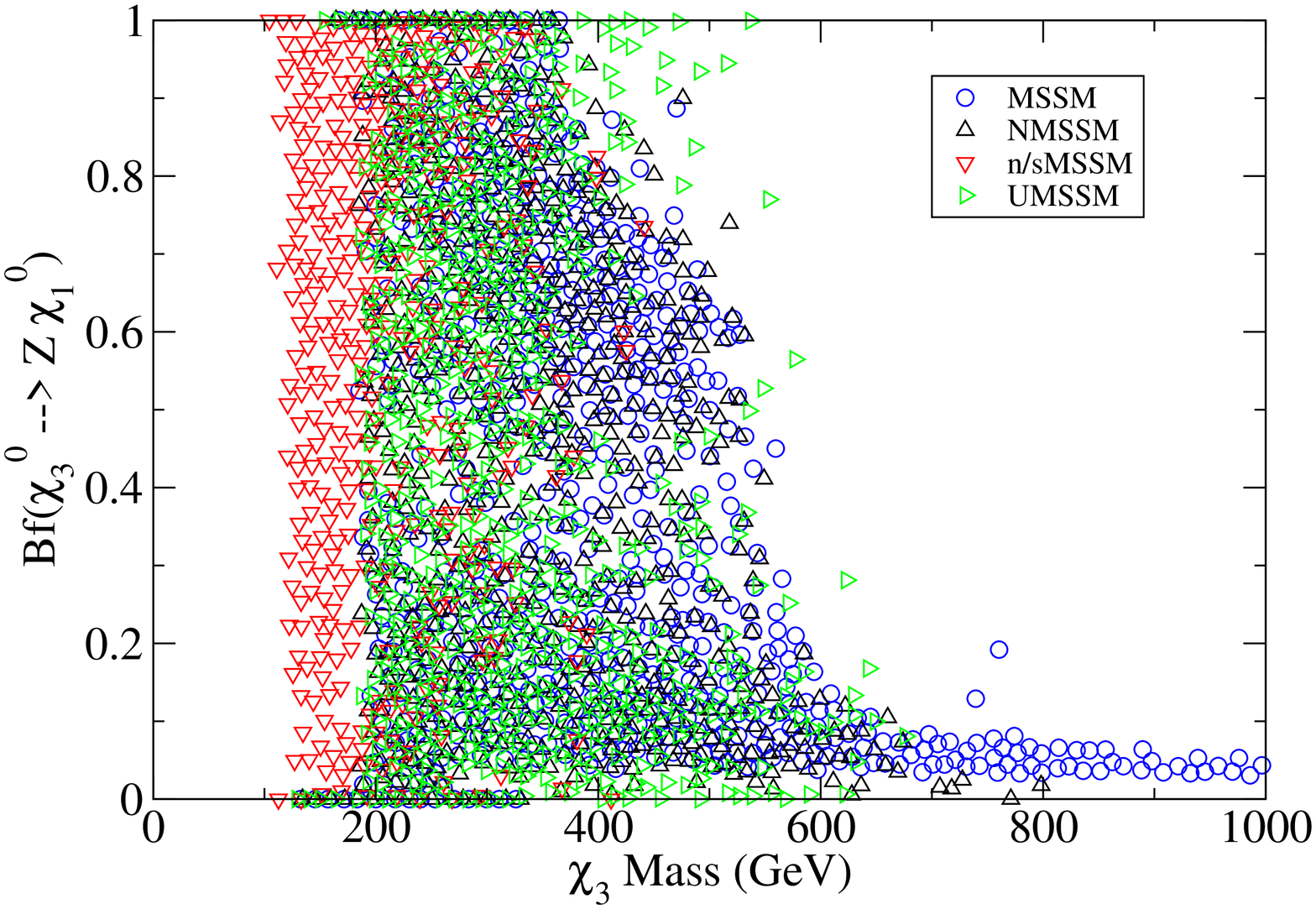}
\includegraphics[angle=-90,width=0.49\textwidth]{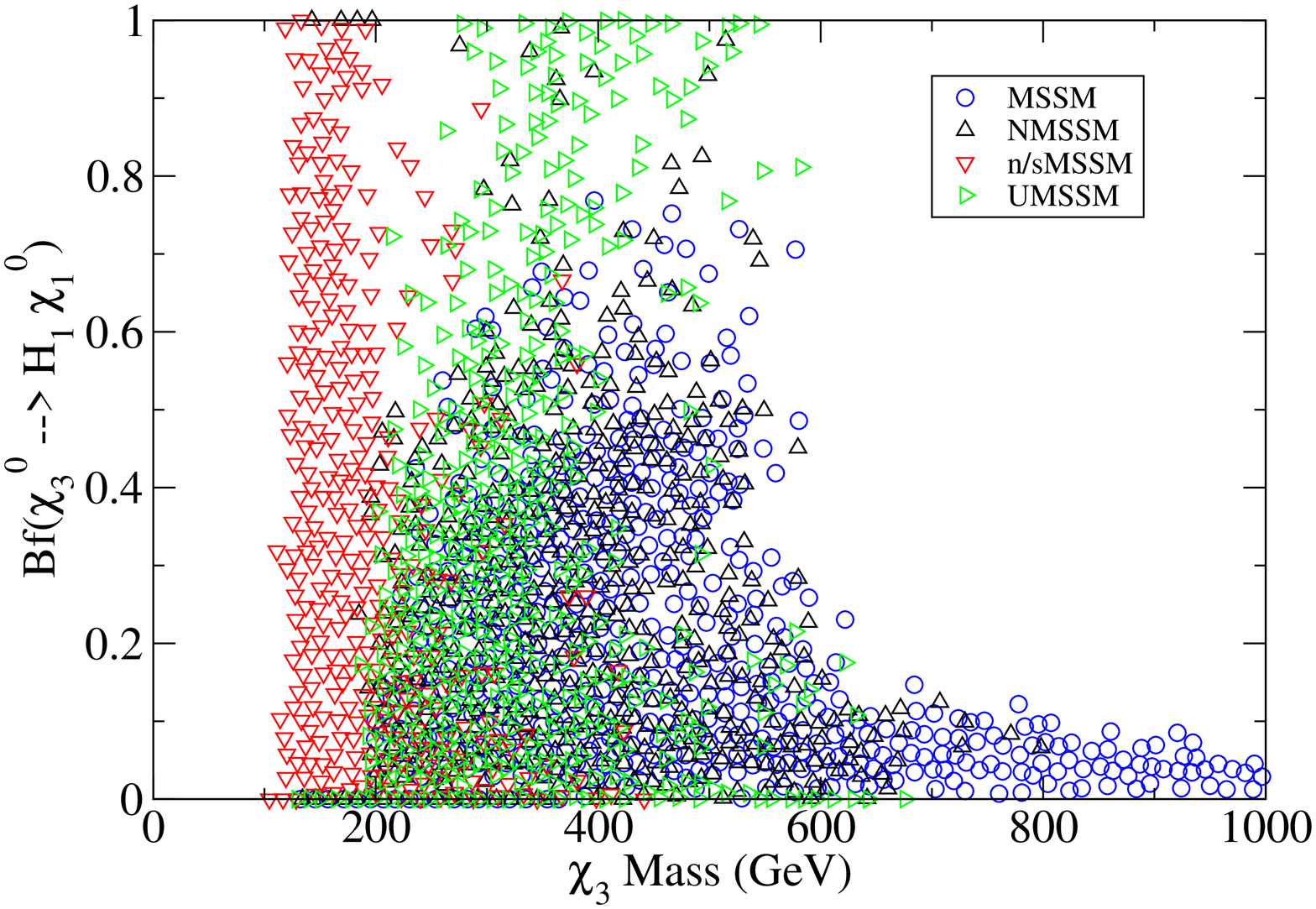}
(e)\hspace{0.48\textwidth}(f)
\caption{Branching fractions of (a) $\N_2 \to \N_1 Z$ and (b)$\N_2 \to \N_1 H_1$ vs. the $\N_2$ mass and  (c) $\N_3 \to \N_2 Z$, (d)$\N_3 \to \N_2 H_1$, (e) $\N_3 \to \N_1 Z$, and (f)$\N_3 \to \N_1 H_1$ vs. the $\N_3$ mass.}
\label{fig:N2-2bdy}
\end{center}
\end{figure}

The two body decays of the second lightest neutralino are typically $\N_2 \to \N_1 Z$ and $\N_1 H_a$.  In Fig. \ref{fig:N2-2bdy}(a,b), we show the branching fractions of $Z$ and $H_1$, respectively.  The mass of $\N_2$ in the n/sMSSM is significantly lighter than in the other models due to the extra neutralino being very light.  The second lightest neutralino $\N_2$ is then similar in mass and composition to $\N_1$ of the MSSM; therefore, an additional step is added in $\N_3$ cascade decays resulting from the extra decay to the very light $\N_1$, e.g.,
\bea
\text{[MSSM]} \quad \N_2 &\to& \N_1 + Z,H_1	\\
\text{[n/sMSSM]} \quad \N_3 &\to& \N_2 + Z,H_1 \to \N_1 + Z,H_1 + Z,H_1
\eea
The extended decay of Eq. 6 is also possible in the other singlet models but less prevalent\footnote{Ellwanger et al. explored neutralino cascades in the NMSSM  with universal
soft supersymmetry breaking terms at $\sim 10^{16}$ GeV for very small values of the couplings $\lambda,\kappa\sim {\cal O}(10^{-2})$ \cite{Ellwanger:1998vi}.}.  This extra decay from an MSSM-like $\N_2$ to a singlino dominated $\N_1$ will be common to SUSY signals, including the end stage of the cascade decays of squarks and gluinos.  In the n/sMSSM, the lightest Higgs can be naturally light and singlet dominated, enhancing the overall rate for that decay chain \cite{Barger:2006dh}.  The decay of an ${\cal O}(100\text{ GeV})$ neutralino into a lighter neutralino and a reconstructed real $Z$ would be characteristic of the n/sMSSM.  Another consequence of the extra light neutralino in the n/sMSSM is a five lepton signal resulting from $\N_2 \C_1$ associated production, which will be discussed further in Section \ref{sect:multilep}.

The branching fractions of $\N_3\to\N_2 Z,H_1$ are also shown in Fig. \ref{fig:N2-2bdy}(c,d); they are typically less than 40\% in most of the models.  The n/sMSSM and UMSSM can have an enhancement due to a singlino (or $Z'$-ino in the UMSSM) dominated lighter neutralino that makes the $\N_3 \to \N_2 X$ decays mimic the $\N_2 \to \N_1 X$ processes in the MSSM ($X=Z,H_1$).  Alternatively, $\N_3$ can decay directly to the lighter $\N_1$ and a $Z$ or Higgs boson with branching fraction shown in Fig. \ref{fig:N2-2bdy}(e,f).  

As evident in Fig. \ref{fig:N2-2bdy}, the neutralino decays in the NMSSM are quite similar to the MSSM since the NMSSM often contains a heavy decoupled singlino.  In the UMSSM, the additional gauge boson, $Z'$, can decay into neutralino pairs \cite{ref:zpdiscovery}, at times producting a heavy $Z'$-ino neutralino.  Heavy neutralino states may also be produced by the cascade decays of heavy gluinos and squarks.

\subsection{Chargino Decays}

Chargino decays are also affected by the singlino via the allowed decay modes to neutralinos.  We calculate the two-body chargino decays 
\be
\C_i \to \C_j W^\pm,~ H^\pm \N_j,~ \C_j Z,~\C_j H_k.
\label{eq:2bdyCdec}
\ee
When these modes are not kinematically accessible, we include the suppressed three body decays $\N_j l \bar \nu_l, \C f \bar f$.  

The most abundantly produced chargino is usually the lightest due to kinematics.  In most of the parameter space the light neutralinos of the NMSSM and UMSSM mimic the MSSM so the chargino decays are not expected to be significantly different from the MSSM.   

The n/sMSSM, with its very light $\N_1$, allows a decay mode of the lightest chargino that is unavailable in the MSSM, namely, the direct decays to the light $\N_1$ and the cascade decay via the $\N_2$:
\be
\C_1 \to \N_2  l \bar \nu \to \N_1 l' \bar l' l \bar \nu.
\ee
This cascade decay contributes to the 5 lepton and 7 lepton signals as discussed further below.

\subsection{Impact on Multilepton events}
\label{sect:multilep}

\begin{figure}[htbp]
\begin{center}
\includegraphics[width=0.79\textwidth]{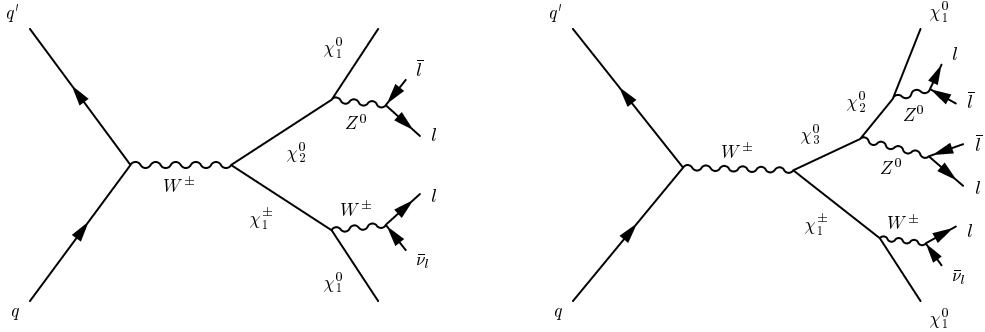}
\caption{Subprocesses that contribute to trilepton (a) and 5$l$ (b) events at hadron colliders.  When the lightest neutralino is dominantly singlino, the trilepton process is enhanced kinematically since $\N_2$ in the xMSSM has the same mass as $\N_1$ in the MSSM with the same parameters.  The 5$l$ events may be associated with $\N_3$ production (or may also be due to $\C_1 \to \N_2 l \bar \nu_l$).}
\label{fig:decay}
\end{center}
\end{figure}

Trilepton events are expected to be one of the prominent signals of supersymmetry at the LHC.  As discussed above, the associated production of the light chargino and neutralino states leads to trilepton events.  In addition, production of $\tilde g \tilde g$, $\tilde l \tilde l$ and $\tilde g \N_{2,3}$ contribute to the trilepton signal if the gluinos and squarks are light \cite{ref:trilep}.  However, if gluinos and sleptons are heavy, most trilepton events will be from chargino and neutralino production via the s-channel $W^\pm$ process.  Some of the contributing neutralino and chargino decays to multileptons are illustrated in Fig. \ref{fig:decay}.

The n/sMSSM yields an enhanced trilepton rate compared to the MSSM as $\N_1$ is lighter, providing more available phase space in the decay.  Further, since the second lightest neutralino in this model is also much lighter than in the MSSM, the production rate of $\N_2$ is enhanced.

\begin{figure}[t]
\begin{center}
\includegraphics[angle=-90,width=0.49\textwidth]{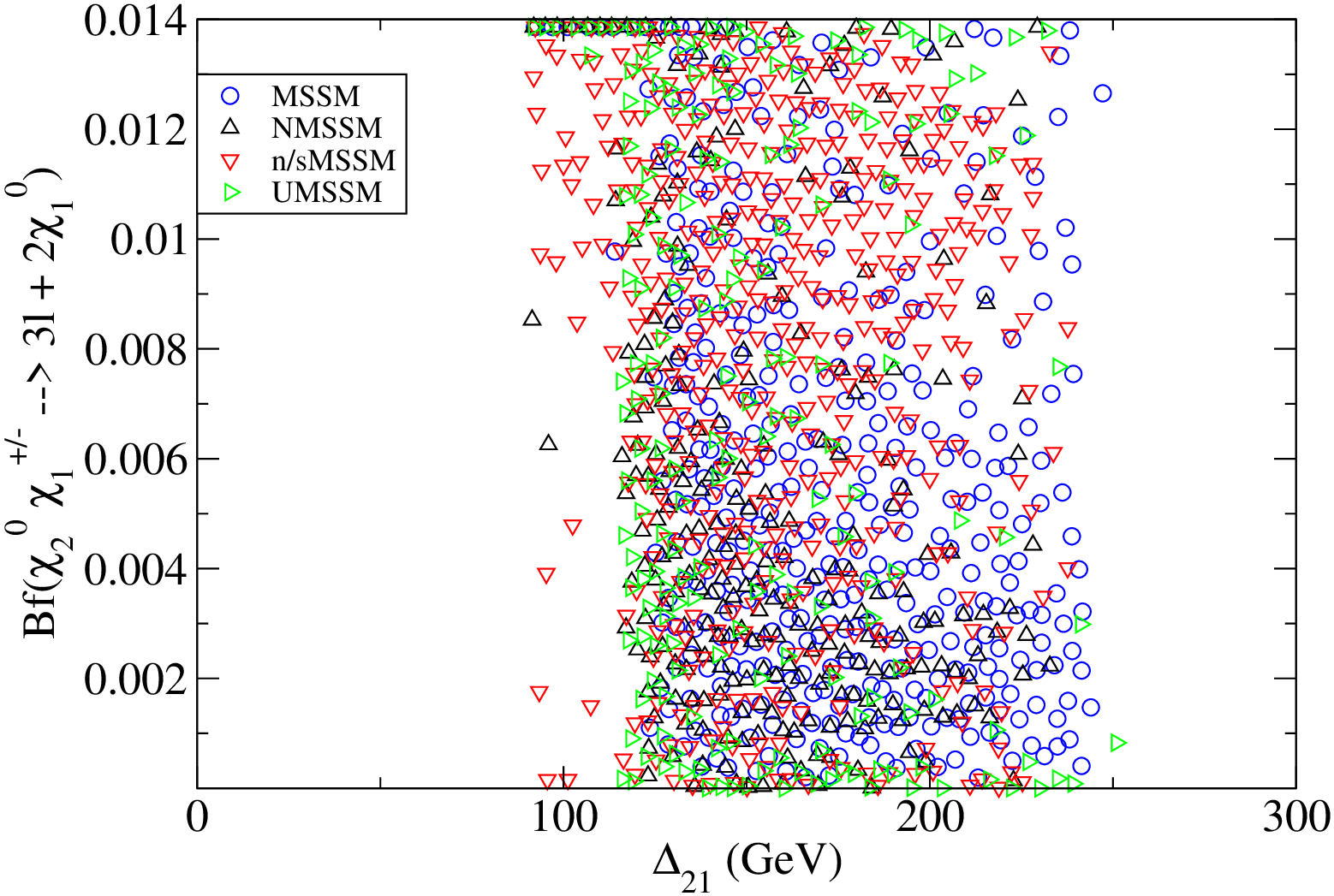}
\includegraphics[angle=-90,width=0.49\textwidth]{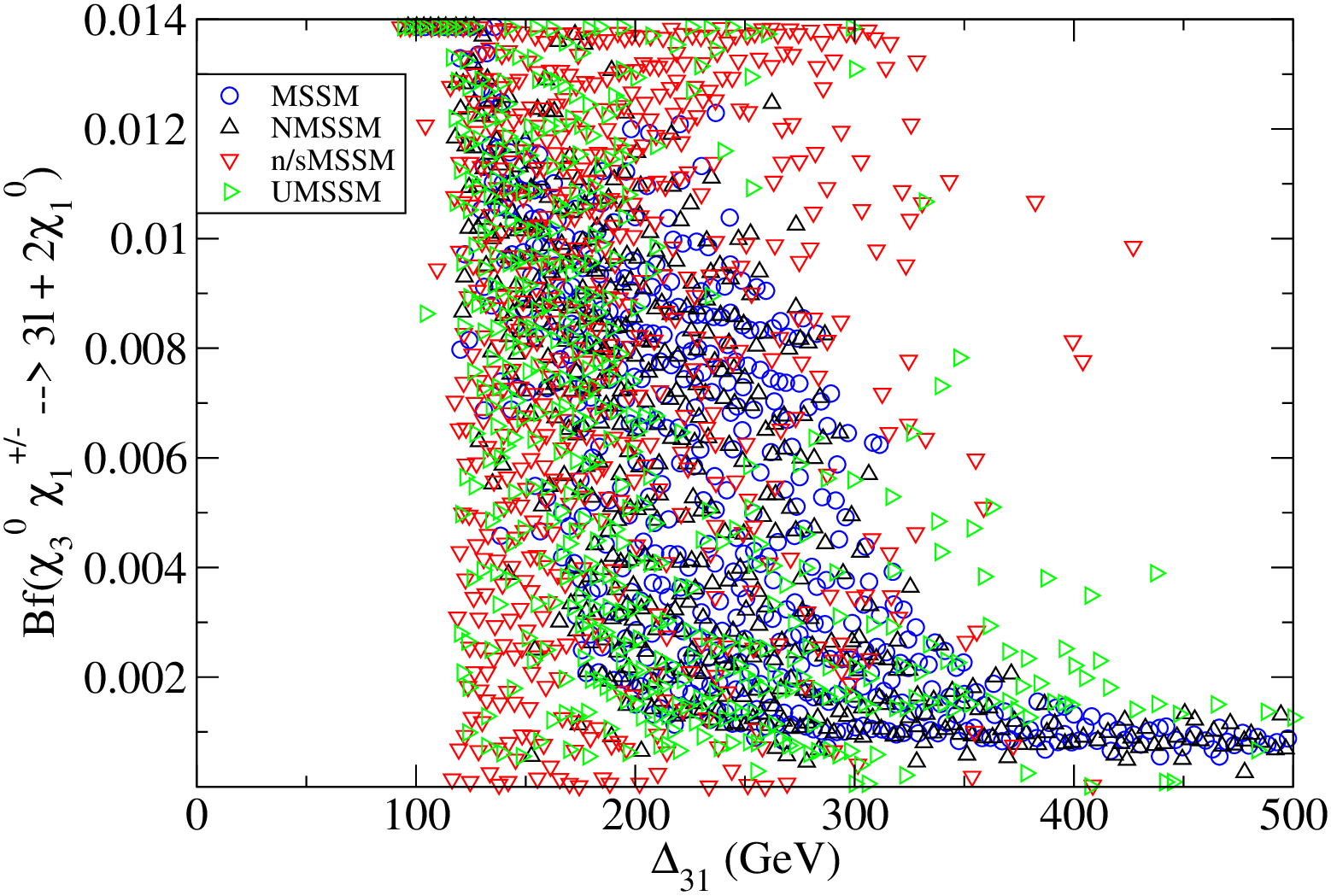}
(a)\hspace{0.49\textwidth}(b)
\includegraphics[angle=-90,width=0.49\textwidth]{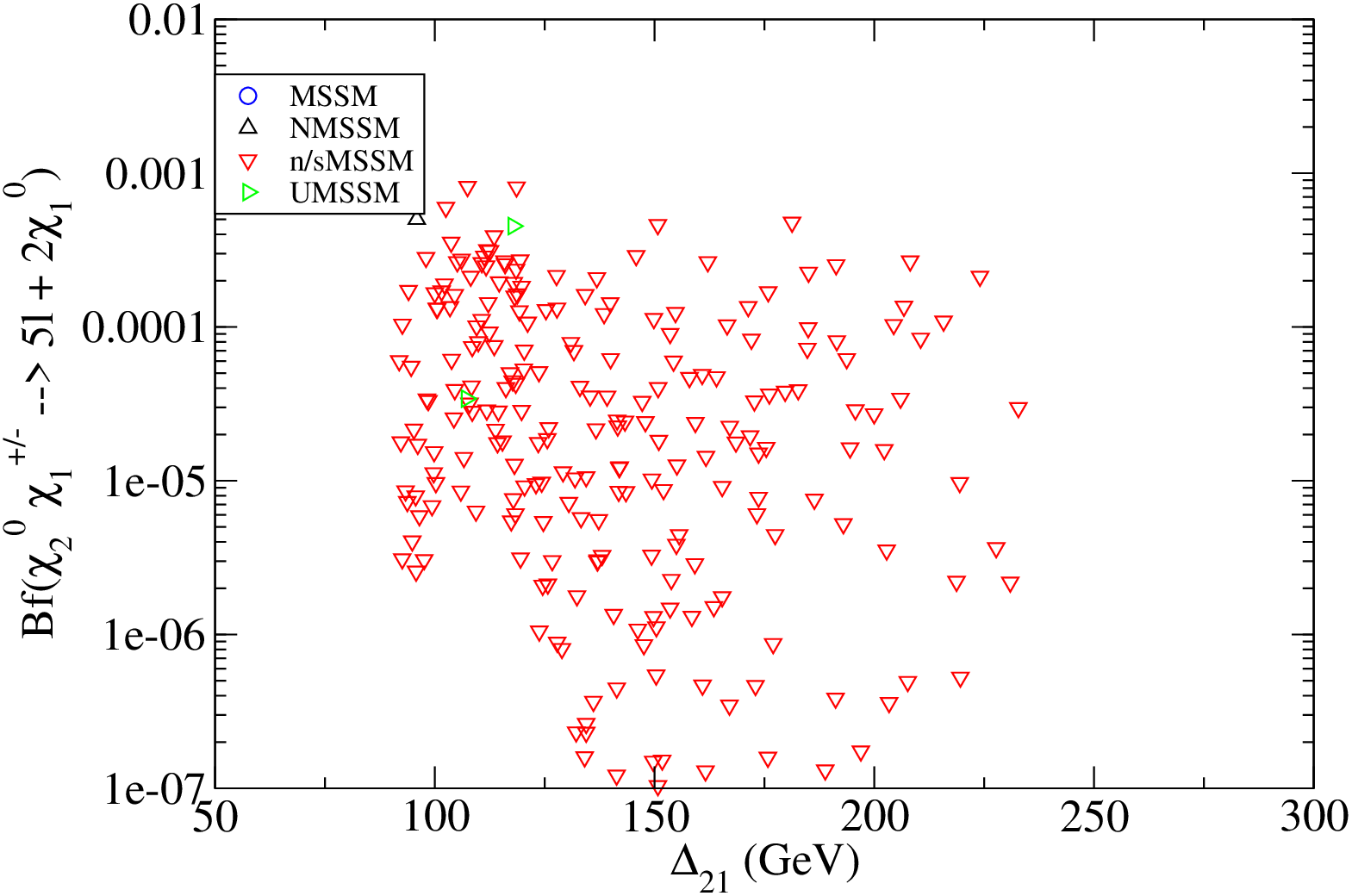}
\includegraphics[angle=-90,width=0.49\textwidth]{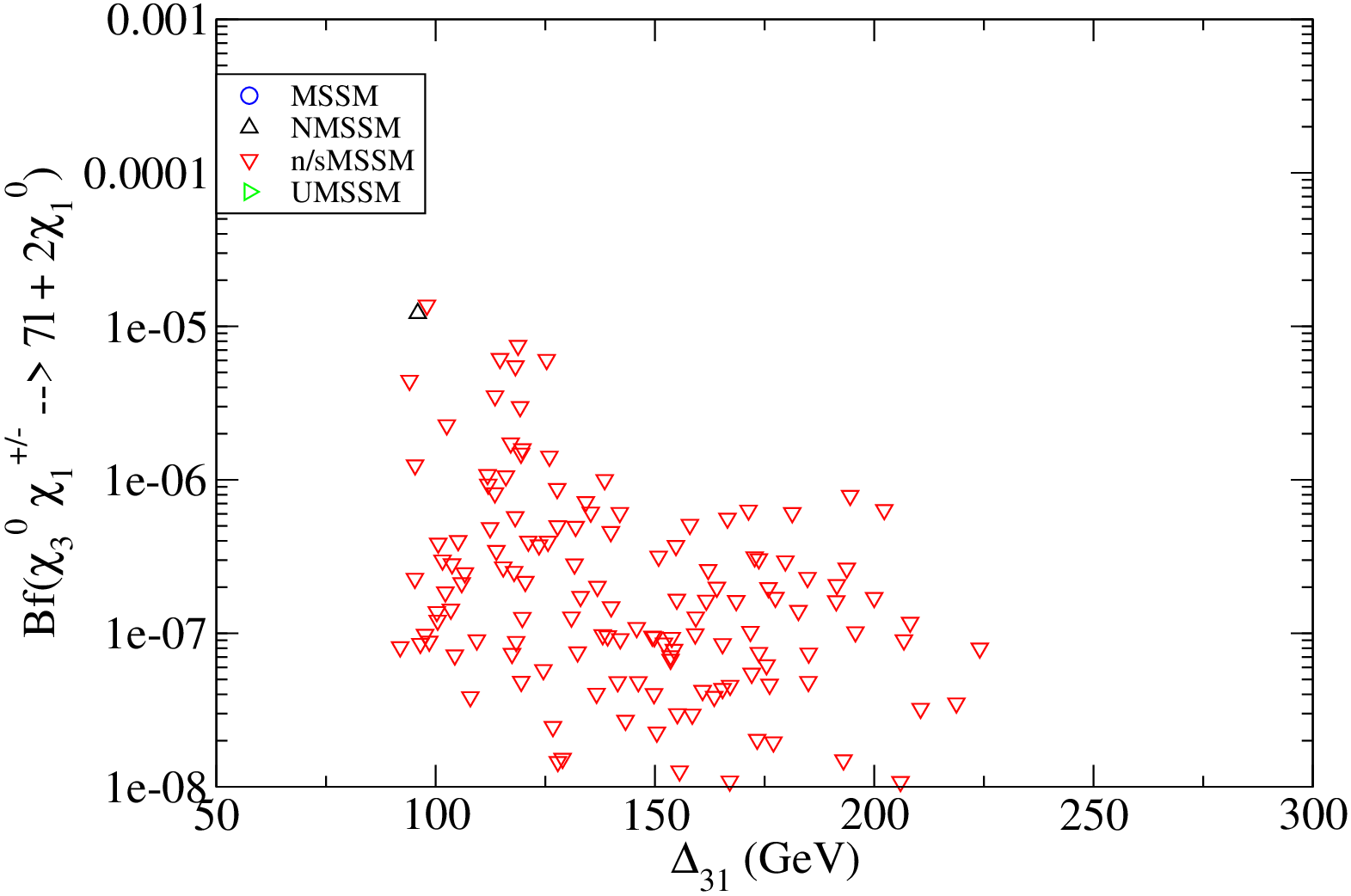}
(c)\hspace{0.49\textwidth}(d)
\caption{Branching ratios of neutralinos and charginos to $3l,5l$ and $7l$ via on shell $Z$ and $W$ bosons vs. the neutralino mass-splitting $\Delta_{ij} = m_{\N_i}-m_{\N_j}$ where the leptons are summed over $e^\pm$ and $\mu^\pm$.  Neutralino decays proceed through (a) $\N_2 \to \N_1 l \bar l$, (b) $\N_3 \to \N_1 l \bar l$, (c,d) $\N_3 \to \N_2 l \bar l \to \N_1 + 4l$.  Chargino decays occur via $\C_1 \to \N_1 W \to \N_1 l \bar \nu_l$ in (a,b) and $\C_1 \to \N_2 W$ in (c,d).  Note that $\C_2 \to \N_2 W$ can  be alternate modes that contribute to a five lepton signal.  Note that the branching fraction of the chargino and neutralino to trileptons in (a,b) are bounded by $Bf(Z \to l^+ l^-) \times Bf(W^\pm \to l \bar \nu) = 0.014$ where $l=e,\mu$.}
\label{fig:trilep}
\end{center}
\end{figure}

Fig. \ref{fig:decay}b shows an example of a xMSSM signal with five leptons in the final state.  In the n/sMSSM, the decay chain starts with $\N_3$, which has about the same mass as the $\N_2$ in the MSSM. Alternatively, in the n/sMSSM 5$l$ can be obtained from extending the chargino decay chain $\N_2 \C_1 \to \N_2 + \N_2 l \bar \nu \to 2 \N_1 + l \bar l + \l \bar l + l \bar \nu$ or from $\N_2 \C_2$ (but this production is kinematically suppressed due to the heavier $\C_2$).  Similar decays from $\N_3 \C_1$ production can result in a $7 l$ final state.  Fig. \ref{fig:trilep} shows predicted branching fraction of produced charginos and neutralinos to trileptons, $5l$ and $7l$ through typical modes via on shell $Z$ and $W$ bosons vs. the mass differences of the neutralinos in the decay.  The branching fraction of the chargino and neutralino to trileptons in (a,b) are bounded by $Bf(Z \to l^+ l^-) \times Bf(W^\pm \to l \bar \nu) = 0.014$ where $l=e,\mu$.The $5l$ and $7l$ events, while rare, usually occur in the n/sMSSM due to the extra step in the decay chain.

The n/sMSSM contains a high density of points in parameter space with relatively high branching fractions to trileptons \footnote{Note that the trilepton production can affect the observed dilepton signals (particularly of interest for like-sign dileptons) if one lepton is not identified.}.  However, the NMSSM trilepton rates are quite similar to the MSSM since the singlino is often decoupled and does not appreciably change the decay kinematics or couplings.

The Higgs sector may also have a strong impact on the rate of multileptons.  The lightest Higgs boson in these models can have masses below the current LEP bound on an MSSM Higgs of 93 GeV, allowing decay modes unavailable in the MSSM.  Since the lightest Higgs of these xMSSM models are typically lighter than 140 GeV, additional pure leptonic contributions are small (except for $\tau^+ \tau^-$, which we are not considering here).  Therefore, models with high branching fractions to the Higgs typically have lower multilepton rates.

As a concrete example, we present the masses, total inclusive associated production cross section of chargino and neutralino pairs through the dominant $s$-channel $W$ process \cite{ref:trilep,Choi:2000kt} and their resulting branching fractions from the cases illustrated in Fig. \ref{fig:level-light} in Table \ref{tbl:illust}.  

\begin{table}[htdp]
\caption{Production cross sections via $s$-channel W and 2-body branching fractions of light neutralinos and charginos in the MSSM and extended models in the scenarios of Fig. \ref{fig:level-light}(a,b).  Branching fractions not shown are kinematically inaccessible.}
\begin{center} 
\begin{tabular}{|c|c||ccc||ccc|}
\hline
	&	&	&	(a) &	&&(b)&\\
\hline
Model	&	MSSM&	N&	n/s &	U&	N&	n/s &	U\\
\hline
$m_{\N_1}$ (GeV)&	120&	120&	3.2 &		120&	119&	18 &	120\\
$m_{\N_2}$ (GeV)&	225&	225&	120 &	225&	217&	121 &	195\\
$m_{\N_3}$ (GeV)&	403&	403&	225 &	398&	292&	228 &	223\\
$m_{\C_1}$ (GeV)&	223&	223&	223 &	223&	223&	223 &	223\\
$m_{H_1}$ (GeV)& 121 &	119&	116&	124 &	131&           115&          119 \\
\hline
$\sigma( pp \to\N_2 \C_1)$ (fb)&	881&	881	&	6.1 	& 880 &780	&	2.8 & 2.3 \\
$\sigma( pp \to\N_3 \C_1)$ (fb)&	4.3&		4.3& 	886	& 4.1&	120& 877 & 872\\
\hline
BF($\N_2\to \N_1 Z$)&	1.00&	0.99&	0.87 &	1.00&	1.00&	1.00 &	--\\
BF($\N_2\to \N_1 H_1$)&	--&		--&	0.13 &	--&		--&	-- &	--\\
\hline
BF($\N_3\to \N_2 Z$)&	0.26&	0.26&	0.12 &	0.26&	--&	0.04 &	--\\
BF($\N_3\to \N_2 H_1$)&	0.04&	0.04&	-- &	0.04&	--&	-- &	--\\
BF($\N_3\to \N_1 Z$)&	0.10&	0.10&	0.50 &	0.11&	0.54&	0.42 &	1.00\\
BF($\N_3\to \N_1 H_1$)&	0.02&	0.02&	0.38 &	0.02&	0.46&	0.53 &	--\\
\hline
BF($\C_1\to \N_1 W$)&	1.00&	1.00&	0.67 &	1.00&	1.00&	0.95 &	1.00\\
BF($\C_1\to \N_2 W$)&	--&	--&	0.33 &	--&	--&	0.05 &	--\\
\hline
\end{tabular}
\end{center}
\label{tbl:illust}
\end{table}%

\subsection{Displaced vertices}

Due to singlino and $Z'$-ino mixing, the neutralinos may have a reduced coupling to SM particles, yielding a suppressed neutralino decay width.  The corresponding decay lengths may be large enough to produce a displaced vertex.  In extreme cases, the $\N_2$ may escape the detector before decay, thereby reducing the number of multilepton events.  There are two causes of a long decay length that we consider: a $\N_2$ that is nearly degenerate with $\N_1$ and a suppressed coupling between $\N_2$ and $\N_1$.

\begin{figure}[h]
\begin{center}
\includegraphics[angle=-90,width=0.49\textwidth]{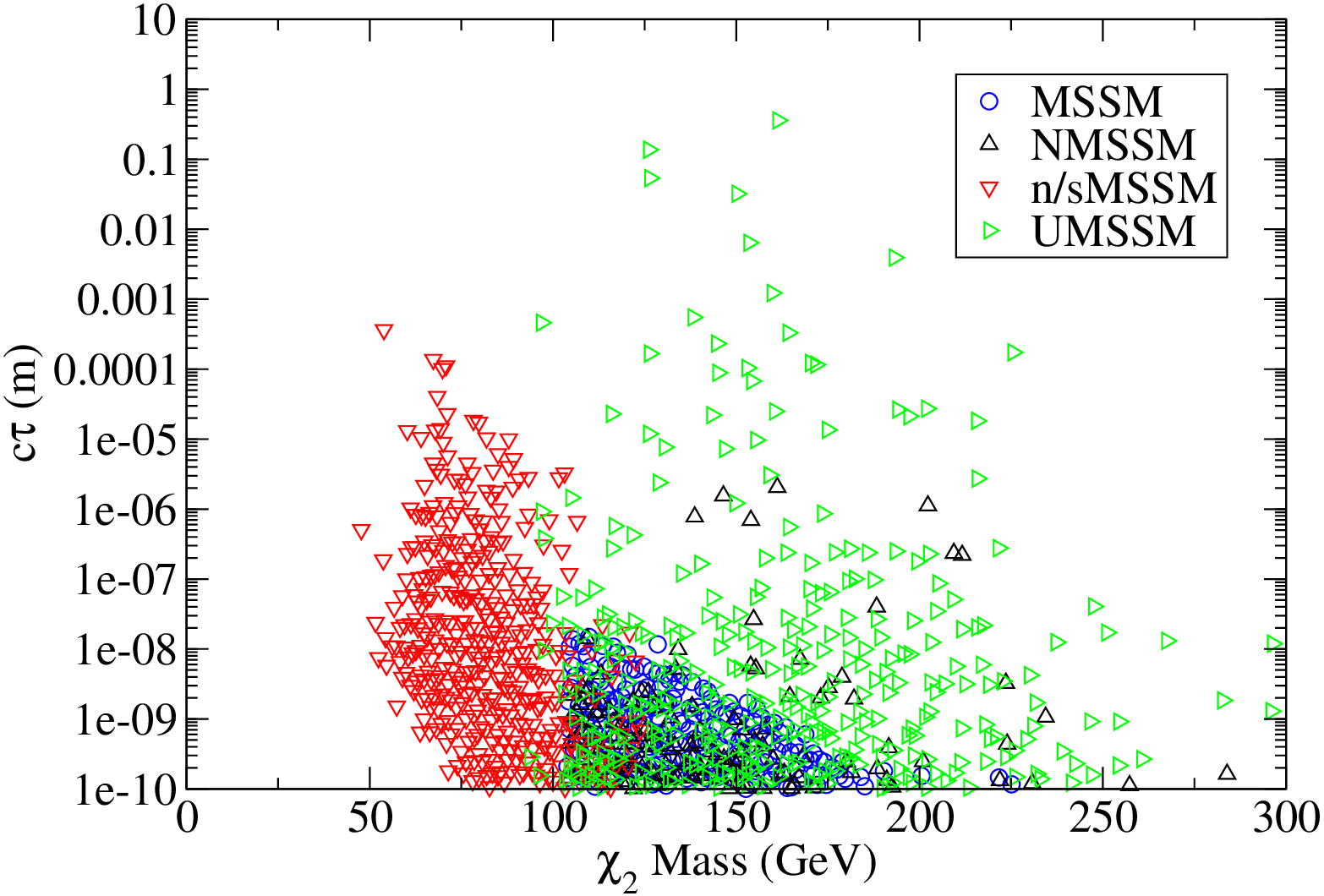}
\includegraphics[angle=-90,width=0.49\textwidth]{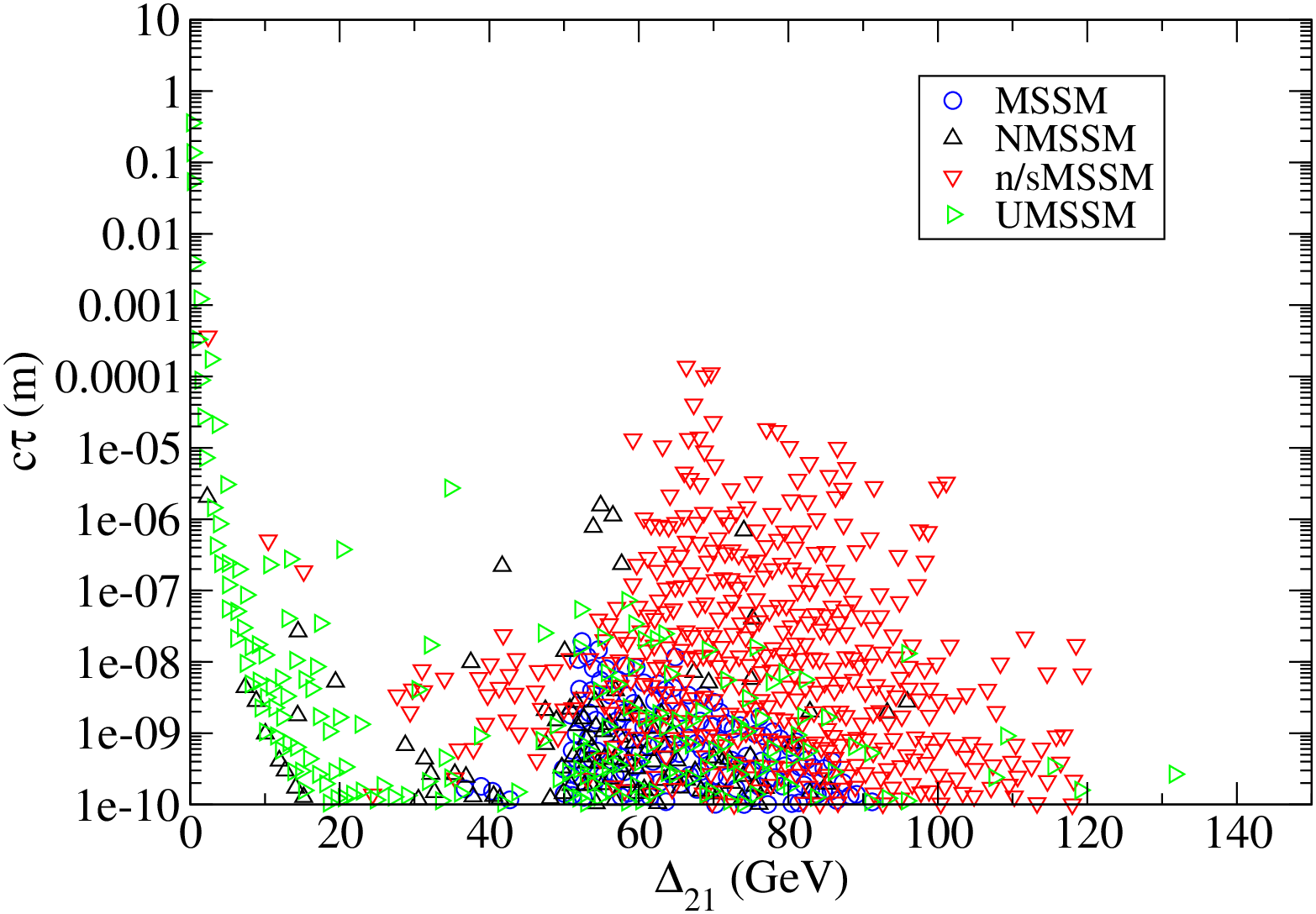}
(a)\hspace{0.49\textwidth}(b)
\caption{ (a) Proper decay length of the $\N_2$ in meters vs. the neutralino mass. (b) Proper decay length of $\N_2$ vs. the mass-splitting, $\Delta_{21} = m_{\N_2}-m_{\N_1}$.}
\label{fig:chi2dl}
\end{center}
\end{figure}

The decay lengths of $\N_2$ are shown in Fig. \ref{fig:chi2dl}a.  A neutralino with typical mass of 100 GeV can have a decay length that may be detectable via a displaced vertex in the UMSSM and n/sMSSM models while the NMSSM and MSSM models yield sub-micron and shorter decay lengths that would be difficult to identify \footnote{Displaced vertices of the NLSP have been explored in the NMSSM in Ref. \cite{Ellwanger:1998vi}}.  The proper decay length can be large in the UMSSM due to kinematic effects of a small mass-splitting of $\N_2$ and $\N_1$.  However, only a small corner of the parameter space allows this accidental degeneracy.  In the n/sMSSM, a long decay length, as long as a few millimeters, is possible due to a small $\N_2 \N_1$ coupling caused by $\N_1$ being dominantly singlino.

If a singlino dominated neutralino is lighter than the lightest chargino, the $\C_1$ decay width can in principle be similarly suppressed.  However, this is far less common with gaugino mass unification, since $\C_1$ can decay into a lighter Bino dominated neutralino.  

\section{Conclusions}\label{sect:concl}

Singlet extended supersymmetric models are known to solve the $\mu$ problem of the MSSM, by the addition of a singlet superfield.  The singlino can affect the neutralino sector considerably, both in the mass spectra and neutralino decays that are expected to give a strong indication of supersymmetry in the form of 3 lepton, 5 lepton or 7 lepton events from cascade decays of neutralinos and charginos.  We emphasize the following properties of the neutralino sector in these singlet extended models:

\bi

\item The extended models can have an approximately decoupled neutralino that is dominantly singlino, accompanied by approximately MSSM neutralino states.  The lightest neutralino is typically very light in the n/sMSSM, often below 50 GeV, whereas the decoupled neutralinos in the NMSSM and UMSSM are heavy.  The neutralinos of the UMSSM and NMSSM can be strongly mixed with the MSSM like neutralinos.  

\item The parameter ranges allowed in the n/sMSSM can be extended beyond those allowed by the MSSM.    A very light neutralino in the n/sMSSM can allow a light stau that is not the LSP, thereby increasing the available parameter regions.

\item The lightest neutralino in the n/sMSSM can significantly impact the predicted numbers of multilepton events.  The $\N_1$ of this model is dominantly singlino and the $\N_2$ is similar in mass to the $\N_1$ of the MSSM if similar model independent parameters are taken.  A new decay chain opens for $\N_3$ ($\sim \N_2$ in the MSSM) that increases the trilepton rate.  An excess in dilepton rates (particularly of interest for like-sign dileptons)  can also occur if one lepton is unidentified.  Moreover, this scenario yields 5 lepton events through the cascade chain $\C_1 + \N_3 \to\N_1 l \bar \nu+ \N_2 2l\to\N_1 l \bar \nu+ \N_1 4l$.  

\item Chargino decays are indirectly affected via their decays to a lighter neutralino state.  The number of neutralino states lighter than the chargino and their respective compositions alter the chargino branching fractions.  This is typically found in the n/sMSSM where the chargino can decay to an MSSM like $\N_2$ and a singlino $\N_1$, yielding another contribution to the 5 lepton signal.  Additionally, the extra step in a chargino decay can allow a 7 lepton final state.  Other models can also exhibit this behavior, but less naturally.

\item The neutralinos of singlet extended models may yield displaced vertices due to macroscopic decay lengths.  If $\N_2$ is either nearly degenerate with the lightest neutralino, or the $\N_2 \N_1$ coupling is suppressed, the proper decay length can be ${\cal O}(1\text{ mm})$ or more.  In the UMSSM, the two lightest neutralinos can be nearly degenerate, resulting in a long decay length.  The $\N_2$ of the n/sMSSM can also have a large decay length as it approximately decouples from the lighter singlino state.  
\ei

Realistic simulations of the LHC signals of extended models will be undertaken in a future study.

\appendix
\section{Neutralino Couplings}
\label{apx:couplings}
The neutralino couplings are altered from their usual MSSM couplings to the Higgs via the additional singlet-singlino (and $Z'$-ino in the UMSSM) interactions, as well as by the extended Higgs sector.  The CP-conserving $H_i \N_j \N_k$ coupling is
\be
C_{\N_i \N_j H_k} = C_{ \N_i \N_j H_k}^L P_L + C_{\N_i \N_j H_k}^R P_R,
\ee
where
\bea
C^R_{\N_i \N_j H_k } &=& \half  \left( ( g_1 N_{i1}-g_2 N_{i2}- g_{1'} Q_{H_d} N_{i6}) N_{j3}+\sqrt 2 \lambda N_{i4} N_{j5} \right) R_{+}^{k1} \nn \\
&+&  \half\left( (g_2 N_{i2}-g_1 N_{i1}- g_{1'} Q_{H_u} N_{i6}) N_{j4}+\sqrt 2 \lambda N_{i3}N_{j5}\right) R_{+}^{k2} \nn \\
&+&  \half\left(\sqrt{2} \lambda N_{i3}N_{j4}-g_{1'} Q_S N_{i6}N_{j5}-\sqrt{2} \kappa N_{i5}N_{j5}\right) R_{+}^{k3}  \nn  \\
&+&  \left( i \leftrightarrow j \right)\\
C^L_{\N_i \N_j H_k } &=& (C^R_{\N_i \N_j H_k })^*.
\eea
where $N_{ij}$ are the rotation matrices which diagonalize the neutralino mass matrix in Eq. \ref{eq:neutmass} and $R_{+}^{ij}$ diagonalize the Higgs mass-squared matrix in the $(H_d, H_u, S)$ basis (see Eq. (26) of Ref \cite{Barger:2006dh}).

\begin{acknowledgments}
This work was supported in part by the U.S.~Department of Energy under grants Nos. DE-FG02-95ER40896 and DOE-EY-76-02-3071 and in part by the Wisconsin Alumni Research Foundation.  We thank H.S. Lee  for helpful discussions and M. McCaskey for helpful checks.  VB thanks the Aspen Center for Physics for hospitality during the course of this work.  
\end{acknowledgments}

\end{document}